\documentclass[12pt]{article}
\usepackage{amsmath}
\usepackage{graphicx}
\usepackage{natbib}
\usepackage{url} 

\usepackage{enumerate}

\usepackage{amssymb}
\usepackage[utf8]{inputenc}
\usepackage{float}
\usepackage{geometry}
\usepackage{caption}
\usepackage{subcaption}

\newcommand{\vect}[1]{\ensuremath{\boldsymbol{\mathbf{#1}}}}	
\newcommand{\matr}[1]{\ensuremath{\boldsymbol{\mathbf{#1}}}}

\newcommand{\sarr}{\sim}
\newcommand{\pdf}[1]{p({#1})}

\newcommand{\pdfp}[2]{p({#1};{#2})}

\newcommand{\pdfc}[2]{p({#1} | {#2} )}

\newcommand{\Gpdf}[4]{\phi_{#1}({#2};{#3},{#4})}

\newcommand{\Sigi}[1]{\boldsymbol{\Sigma}_{#1}}
\newcommand{\sigi}[1]{\sigma_{#1}^2}

\newcommand{\Expe}[1]{\boldsymbol{E} [ {#1} ] }

\newcommand{\Vari}[1]{\boldsymbol{Var} [ {#1} ]}

\newcommand{\iv}[1]{\boldsymbol{i}_{#1}}
\newcommand{\Idm}[1]{\boldsymbol{I}_{#1}}

\newtheorem{algorithm}{Algorithm}

\newcommand{\blind}{0}

\begin{document}

\def\spacingset#1{\renewcommand{\baselinestretch}%
{#1}\small\normalsize} \spacingset{1}


\if0\blind
{
  \title{\bf  The Spatial Kernel Predictor based on Huge Observation Sets}
  \author{Henning Omre \\
  	Department of Mathematical Sciences, Norwegian University of Science \& Technology\\
  	and \\
  	Mina Spremi\'c\\
  	Department of Mathematical Sciences, Norwegian University of Science \& Technology}
  \date{}
  \maketitle
} \fi

\if1\blind
{
  \bigskip
  \bigskip
  \bigskip
  \begin{center}
    {\LARGE\bf Title}
\end{center}
  \medskip
} \fi

\bigskip
\begin{abstract}
For a Gaussian random field model the Kriging predictor coinsides with the conditional expectation given the observation set. 
The spatial predictor is usually the Kriging predictions, with prediction variances, in the nodes of a relatively dense regular grid over the study domain.
An alternative expression for this conditional expectation provides a spatial predictor on functional form which does not rely on a spatial grid discretization. This functional predictor, called the Kernel predictor, is identical to the asymptotic grid infill limit of the Kriging-based grid representation. The computational demand is primarily dependent on the number of observations - neither on any spatial grid discretization nor on the dimension of the spatial reference domain.

We explore the potential of this Kernel predictor with associated prediction variances. The predictor is valid for Gaussian random fields with any eligible spatial correlation function, and large computational savings can be obtained by using a finite-range spatial correlation function. For studies with a huge set of observations, localized predictors must be used, and the computational advantage relative to Kriging predictors can be very large. Moreover, model parameter inference based on a huge observation set can be efficiently made. The methodology is demonstrated on two examples. 

\end{abstract}

\noindent%
\vfill

\newpage
\spacingset{1} 
\section{Introduction}
\label{sec:intro}
Many applications involve spatial prediction either as the ultimate goal or as an intermediate effort. Prediction of a spatial variable $ \{ r ( \vect{x} ) ; \vect{x} \in \texttt{D} \} $ based on a set of $ m $ observations 
$ \vect{ r}^d = 
( r ( \vect{x}_1^d ), \dots , r ( \vect{x}_m^d ) ) ; \vect{x}_i^d \in \texttt{D} $ is often the challenge, but frequently also more complicated observation acquisition procedures must be modeled. 
The prediction of the spatial variable in an arbitrary location $ \vect{x}_+ \in \texttt{D} $
is usually performed by Kriging, see \cite{Chiles2012}, which is optimal under a Gaussian Random Field (RF) model. If the entire spatial variable over $ \texttt{D} $ is of interest, a grid representation is often used, where Kriging predictions and prediction variances are assigned 
to each grid node. The grid is usually defined sufficiently dense such that predictions at non-grid locations are made by simple piecewise constant or piecewise planar interpolation. Particularly in studies with a huge number of observations, the grid needs to be dense to capture the information content in the observations. Hence only predictions and prediction variances in each grid node are computed and stored, and one needs not solve the entire Kriging system to assess the spatial variable in an arbitrary location. This grid representation can also be the base for contour or perspective presentations, and further processing of the predicted spatial variable with associated prediction variances.

In \cite{Heaton2018} an overview and an evaluation of various spatial predictors suitable for studies with large observation sets, hence with $ m $ large, are presented.
Since also the grid dimension $ n $ usually is high in these studies the computational demands are often large. The Gauss Markov RF model, see \cite{GMRFbook}, is defined on the actual grid system 
and by specifying a high degree of conditional independence in the grid values, these computational demands are reduced. In cases with locationwise, exact observations in a subset of grid nodes, large computational savings can be obtained. If the observations are collected in a more general manner, the predictor is on the contrary, very inefficient. For these cases the calculations in the predictor involve matrices defined by the grid dimension.

Spatial variables can also be represented on functional form, in a basis function representation,
see \cite{Buhmann2003} and \cite{Cressie2022}. The traditional Fourier representation is discussed in \cite{Borgman1984} and other finite-support basis functions are used in \cite{Cressie2008} and \cite{Nychka2015}. The former is not well suited for conditioning on observations, while the latter relies on the spatial references of the basis functions and hence an underlying grid representation must be defined. Also the Stochastic Partial Differential Equation model, see \cite{Lindgren2011}, uses a functional representation, but the associated spatial predictor requires an underlying grid system to be defined. In all the predictors relying on an underlying grid representation, operations on matrices defined by the grid dimension $ n $ are involved in the conditioning procedure.

The Kriging predictor is identical to the conditional expectation given the observations under the Gaussian RF model assumption, and the expression only involves a matrix defined by the dimension of the observation set $ m $. The same holds for the expression for the prediction variance. 
Hence the prediction of the spatial variable in one location only requires inversion of a matrix of dimension $ (m \times m) $, but for a grid representation the prediction must be repeated in all $ n $ grid nodes. 

We define a predictor for the spatial variable represented on functional form, with associated prediction variances. 
This predictor is identical to the Dual Kriging predictor discussed in \cite{Matheron1971} and \cite{Galli1984}, but we present an alternative development of the predictor and extend its use to studies with huge observation sets, both for prediction and model parameter inference.
We refer to the predictor as the Kernel predictor, and it is, as the Kriging predictor, optimal under Gaussian RF model assumptions. The Kernel predictor actually appears as the grid infill asymptotic limit of a grid based Kriging predictor. The predictor only involves operations on matrices defined by the observation dimension $ m $, no underlying grid representation is required, which may provide large computational advantages.

In order to improve computational efficiency even more, we consider a class of finite-range spatial correlation function models, see \cite{Gneiting2002}. Under these model assumptions the observation inter-correlation matrix appears as sparse, for which efficient Cholesky decomposition algorithms are available.

In studies with huge amounts of observations, localized approximations of the predictor must be used, see \cite{Chiles2012}. We define a localized approximate Kernel predictor which is far more computer efficient than the traditional localized Kriging predictor.
Model parameter inference is also defined for this case.

In the notation, greek letters represent model parameters. Vectors are denoted by bold lower case letters while matrices are denoted by bold upper case ones and upper-index $ T $ entails transpose. A random Gaussian 
$ n $-vector $ \vect{r} $  with expectation 
$ \Expe{ \vect{r} } = \vect{ \mu }_r $ and covariance $ \Vari{ \vect{r } } = \Sigi{r} $
has a probability density function (pdf) represented by 
$ \Gpdf{n}{ \vect{r} }{ \vect{\mu}_r }{ \Sigi{r} } $.
The $ n $-vector $ \iv{n} $ is a vector with entries of ones only.

\section{Model Definition}

Consider a continuous spatial variable  $ \{ r ( \vect{x} ) ; \vect{x} \in \texttt{D} \subset \mathbb{R}^q \}; r ( \vect{x} ) \in \mathbb{R} $ with spatial reference variable $ \vect{x} $ in a bounded $ q $-dimensional domain $ \texttt{D} $, with $ q  = 1, 2 $ or $ 3 $.
Let the spatial variable be observed in $ m $ locations represented by the spatial reference set 
$ \texttt{M}: \{ \vect{x}_1^d , \dots ,\vect{x}_m^d \} ; \vect{ x }_i^d \in \texttt{D} $, also refered to as observation number $ i = 1, \dots , m $. 
The corresponding observed values are represented by the $ m $-vector 
$ \vect{ r}^d = ( r ( \vect{x}_1^d ), \dots , r ( \vect{x}_m^d ) )^T 
= \{ r_{ \vect{y} }^d ; \vect{y} \in \texttt{M} \} $.
The observations are assumed to be locationwise and exact only to simplify notation, the developments can easily be extended to observations as any linear operator on the spatial variable with additive Gaussian error, as discussed later in the paper.

The challenge is to assess the spatial variable given the observation $ m $-vector $ \vect{ r }^d $, hence,
$ \{ [ r ( \vect{x} ) | \vect{r}^d ] ;\\ \vect{x} \in \texttt{D}  \} $. 
Note in particular that a spatial representation of the conditional spatial variable is the objective.

Assign a prior probabilistic model to the spatial variable of interest 
$ \{ r ( \vect{x} ) ; \vect{x} \in \texttt{D} \} $. Assume that it is a stationary Gaussian RF with expectation and variance levels $ \mu_r \in \mathbb{R} $ and $ \sigi{r} \in \mathbb{R}_\oplus $, and a spatial correlation function 
$ \rho_r ( \vect{\tau} ) \in \mathbb{R}_{ [ -1,1 ] } ; \vect{\tau} =  \vect{x}'' - \vect{x}'  $, being a non-negative definite function. 
Assume further that $ \rho_r ( \vect{ \tau} ) $ is continuous at the origin without 'nugget' effect, hence the spatial variable is continuous almost everywhere. The developments can easily be extended to cover cases with spatial correlation functions with 'nugget' effects.

Consider an arbitrary location $ \vect{x}_+ \in \texttt{D} $ in which the value of the conditional spatial variable shall be assessed. Then from the Gaussian RF assumptions above, we have,
\begin{align}  
	\left[
	\begin{array}{c}
		r (\vect{x}_+) \\  \vect{r}^d
	\end{array} \right]
	\sarr
	\pdf{r_+,\vect{r}^d}
	= \phi_{1+m} \left( 
	\left[
	\begin{array}{c}
		r_+ \\  \vect{r}^d
	\end{array} \right]
	; \left[ \begin{array}{c} 
		\mu_r  \\  \mu_r   \iv{m}
	\end{array}  \right] , 
	\left[
	\begin{array}{cc}
		\sigi{r}   &   \vect{\sigma}_{+d}  \\
		\vect{\sigma}_{d+} &   \Sigi{d} 
	\end{array}  \right] 
	\right)  \nonumber
\end{align}
where $ \Sigi{d} = \sigi{r}  \Sigi{d}^\rho $ and
$ \vect{\sigma}_{d+} = \sigi{r}  \vect{\rho}_{d+} $ 
with the correlation $ ( m \times m)$-matrix $ \Sigi{d}^\rho $ containing the inter-correlations between the observations, hence 
$ \rho_r ( \vect{ \tau}_{ij}^d ) ; \vect{ \tau }_{ij}^d = \vect{x}_i^d - \vect{x}_j^d 
; i,j = 1, \dots , m $. The correlation $ m$-vector $ \vect{ \rho}_{d+} $ contains the corresponding correlations between the observations and the spatial variable value to be predicted; 
$ \rho_r ( \vect{ \tau }_{i+}^d) ; \vect{ \tau}_{i+}^d = \vect{x}_i^d - \vect{x}_+;i =1, \dots,m $, and $ \vect{ \rho}_{+d} =  \vect{ \rho}_{d+}^T  $.

It follows from familiar Gaussian theory that,
\begin{align*}
	[ r ( \vect{x}_+) | \vect{r}^d ] \sarr \pdfc{ r_+ }{ \vect{r}^d } 
	= \Gpdf{1}{ r_+}{ \mu_{ + | d }}{ \sigi{ + | d } }
\end{align*}
where,
\begin{align*}
	\Expe{ r ( \vect{x}_+ ) | \vect{r}^d } &= \mu_{+|d} 
	= \mu_r + \vect{\sigma}_{+d} [  \Sigi{d} ]^{-1} ( \vect{r}^d - \mu_r \iv{m} ) 
	= \mu_r + \vect{\rho}_{+d} [  \Sigi{d}^\rho ]^{-1} ( \vect{r}^d - \mu_r \iv{m} ) \\
	\Vari{ r ( \vect{x}_+ | \vect{r}^d } &= \sigi{+|d}
	= \sigi{r} - \vect{\sigma}_{+d} [  \Sigi{d} ]^{-1} \vect{\sigma}_{d+} 
	= \sigi{r} [ 1  - \vect{\rho}_{+d} [  \Sigi{d}^\rho ]^{-1} \vect{\rho}_{d+} ]
\end{align*}
Based on these relations one obtains the predictor and predictor variance in the arbitrary location $ \vect{x}_+ \in \texttt{D} $,
\begin{align} \label{eq:GRF-Pred}
	\hat{r}_+ &= \mu_{+|d} 
	=  \mu_r + \vect{\rho}_{+d} [  \Sigi{d}^\rho ]^{-1} ( \vect{r}^d - \mu_r \iv{m} ) \\
	\sigi{+} &= \sigi{ + | d} 
	=  \sigi{r} [ 1   -  \vect{\rho}_{+d} [  \Sigi{d}^\rho ]^{-1} \vect{\rho}_{d+} ]
	\nonumber
\end{align}
Under the current model assumptions these expressions are exact and optimal in both a maximum posterior and squared error sense, given the model parameters 
$ [ \mu_r, \sigi{r} , \rho_r ( \vect{ \tau} ) ] $.

\vspace{5 mm}
\noindent
\textbf{Model Parameter Inference }  \label{sec:GRF-Inf}

The parameters of the model are expectation level $ \mu_r \in \mathbb{R} $, 
variance level $ \sigi{r} \in \mathbb{R}_\oplus $ and 
the spatial correlation function $ \rho_r ( \vect{ \tau } ; \vect{\eta}_r ) $ parametrized by 
$ \vect{ \eta}_r \in \mathbb{R}^\kappa $. These model parameters may be assessed by a maximum marginal likelihood criterium,
\begin{align*}
	( \hat{ \mu}_r , \hat{ \sigma }_r^2 , \hat{ \vect{\eta}}_r ) 
	&= \arg \max_{\mu_r , \sigi{r} , \vect{ \eta}_r } 
	\{  \pdfp{\vect{r}^d }{ \mu_r , \sigi{r} , \vect{ \eta}_r } \} \\
	&= \arg \max_{\mu_r , \sigi{r} , \vect{ \eta}_r } 
	\{ [ 2 \pi ]^{-m/2} [ \sigi{r} ]^{-m/2} | \Sigi{d}^{\rho|\eta_r } |^{-1/2} \\
	&\times \exp \{ -  [ 2 \sigi{r} ]^{-1} ( \vect{r}^d - \mu_r \iv{m} )^T 
	[ \Sigi{d}^{\rho | \eta_r } ]^{-1} ( \vect{r}^d - \mu_r \iv{m} ) \}
\end{align*}
Note that the three maximum conditional marginal likelihood estimators are,
\begin{align} \label{eq:MLE-ExpVar}
	( \hat{ \mu}_r | \sigi{r} , \vect{\eta}_r ) 
	&= [ \iv{m}^T [ \Sigi{d}^{\rho | \eta_r} ]^{-1} \iv{m} ]^{-1} 
	\times \iv{m}^T [ \Sigi{d}^{\rho | \eta_r} ]^{-1} \vect{r}^d \\
	( \hat{ \sigma}_r^2 | \mu_r , \vect{ \eta}_r )
	&= m^{-1} ( \vect{r}^d - \mu_r \iv{m} )^T [ \Sigi{d}^{\rho | \eta_r} ]^{-1} 
	( \vect{r}^d - \mu_r \iv{m} )  \nonumber \\
	( \hat{\vect{ \eta}}_r | \mu_r , \sigi{r} ) 
	&= \arg \min_{ \vect{ \eta}_r } 
	\{ \ln | \Sigi{d}^{ \rho | \eta_r } | + 
	( \vect{r}^d - \mu_r \iv{m} )^T [ \sigi{r} \Sigi{d}^{\rho | \eta_r} ]^{-1} 
	( \vect{r}^d - \mu_r \iv{m} ) \}.  \nonumber
\end{align}
The two former optimizations are sequentially analytically tractable, hence 
$ [ \hat{\mu}_r , \hat{ \sigma}_r | \vect{ \eta}_r ] $ can be assessed analytically. The latter optimization must be made numerically, but since the 
parameter $ \kappa $-vector $ \vect{\eta}_r $ usually is low-dimensional this optimization normally is feasible. Hence the maximum marginal likelihood 
estimator  $ [ \hat{\mu}_r , \hat{ \sigma}_r , \hat{ \vect{ \eta} }_r ] $ can be assessed iteratively.

The model parameters in the spatial correlation function $ \vect{ \eta }_r $ are notoriously complicated to estimate reliably, since the short-distance correlation is most important and the observations are usually some distance apart. The range parameter can often be inferred by visual inspection of non-parametric correlation estimates. Model parameter inference is not in focus of this study, hence we assume that $ \vect{ \eta }_r $ can be assessed informally. Then the estimates of the expectation and variance levels 
$ [ \hat{\mu}_r , \hat{ \sigma}_r | \vect{ \eta}_r ] $ can be calculated analytically if
the $ ( m \times m ) $-matrix
$  [ \Sigi{d}^{\rho | \eta_r} ]^{ -1}  $ can be calculated.

The model assumptions specified above can be extended to spatial regression with the expectation function being $ \mu( \vect{x}) = \vect{ g}_{\vect{x}}^T \vect{ \beta}_r $ 
where the $ ( L+1)$-vector $ \vect{ g}_{\vect{x}} = [ 1, g_1( \vect{x} ), \dots , g_L( \vect{x} )]^T $ contains $ L $ known regression functions for $ \vect{x} \in \texttt{D} $, while
the $ ( L + 1) $-vector $ \vect{ \beta }_r = [ \beta_{r0}, \dots , \beta_{rL} ]^T $ contains the associated unknown regression coefficients. Note that the first element represent the expectation level. Maximum conditional likelihood expressions similar to the ones given above can be developed for this spatial regression case, on the expense of more complex notation.

\section{Spatial Predictors} \label{sec:Spat-Pred}

The objective is to provide a predictor of the spatial variable 
$ \{ r( \vect{x} ) ; \vect{x} \in \texttt{D}  \} $
based on the observation set $ \vect{r}^d $, with associated predictor variances. The predictor for the value of the spatial variable at an arbitrary location $ \vect{x}_+ \in \texttt{D} $, specified in Expression \ref{eq:GRF-Pred}, must be activated as $ \vect{x}_+ $ slides across the spatial domain $ \texttt{D} $. There are two alternative representations of the spatial prediction, either a spatially discretized grid representation or a spatially continuous functional representation. 

\vspace{5 mm}
\noindent
\textbf{The Traditional Kriging Predictor}

The traditional Kriging predictor, see \cite{Chiles2012}, being the best linear unbiased 
predictor (BLUP) in the squared-error sense, coinsides with the conditional expectation under Gaussian RF model assumptions. The spatial predictor is usually on a grid representation in the $ n $ nodes of a regular grid $ \texttt{L} $ covering the reference domain $ \texttt{D} $.
The grid $ \texttt{L} \subset \texttt{D} $ must be dense relative to the observation density in order to capture variability caused by the conditioning, hence $ n \gg m $.
From Expression \ref{eq:GRF-Pred} one has,
\begin{align*}
	\{ \hat{r} ( \vect{x} )  = \mu_r + \vect{\alpha}^{xT} ( \vect{r}^d - \mu_r \iv{m} )
	= \mu_r + 
	\Sigma_{ \vect{y} \in \texttt{M} } \alpha_{ \vect{y} }^x ( r_{ \vect{y} }^d - \mu_r );
	\vect{x} \in \texttt{L} \subset \texttt{D}  \}
\end{align*}
with Kriging-weight $ m $-vector 
$ \vect{ \alpha }^x = [ \Sigi{d} ]^{-1} \vect{ \sigma }_{dx} 
= [ \Sigi{d}^\rho ]^{-1} \vect{ \rho }_{dx} 
= ( \alpha_1^x, \dots , \alpha_m^x )^T $. Hence the predictor appears as a linear combination of the observed values $ \vect{r}^d $. The corresponding prediction variance expression is,
\begin{align*}
	\{ \sigi{p} ( \vect{x} ) = \sigi{r} 
	- \vect{\alpha}^{xT} \Sigi{d} \vect{ \alpha}^x 
	; \vect{x} \in \texttt{L} \subset \texttt{D}  \}.
\end{align*}
The Kriging-weight $ m $-vector $ \vect{ \alpha}^x $ is dependent on the prediction location $ \vect{x} \in \texttt{D}  $ and the observation locations $ \texttt{M} $, but independent of the actual values of the observations $ \vect{r}^d $. Hence these weights are dependent on the model parameters only, and therefore deterministic. 
The $ ( \vect{r}^d - \mu_r \iv{m} ) $-term is random, however, with
$ \Expe{ \vect{r}^d - \mu_r \iv{m} } = 0  \iv{m} $ and 
$ \Vari{ \vect{r}^d - \mu_r \iv{m} } = \Sigi{d} $.

The weight $ m $-vector $ \vect{ \alpha}^x = [ \Sigi{d}^\rho ]^{-1} \vect{\rho}_{dx} $
must be calculated for each grid node, hence $ n $ times. These calculations need not be too computer demanding however, since the computer demanding term 
$ [ \Sigi{d}^\rho]^{-1} $ is global and common in all weight calculations and therefore only needs to be calculated once.
The spatial predictor is based on this grid representation and it is usually defined as either piecewise constant or piecewise planar. The former predictor, at an arbitrary location, takes the value of the closest grid node. The latter predictor, at an arbitrary location, is based on a linear interpolator of values in the closest grid nodes.

\vspace{5 mm}
\noindent
\textbf{The Kernel Predictor}

The Kernel predictor introduced in this study has a form similar to the Dual Kriging predictor, see \cite{Galli1984} and \cite{Chiles2012}. It appears by an alternative phrasing of the conditional expectation under Gaussian RF model assumptions. The spatial predictor is on a functional representation. From Expression \ref{eq:GRF-Pred} one has,
\begin{align} \label{eq:KP-Pred}
	\{ \hat{r} ( \vect{x} ) 
	= \mu_r + \vect{ \rho }_{xd} \vect{ \alpha }^d 
	= \mu_r + \Sigma_{ \vect{y} \in \texttt{M} } 
	\alpha_{ \vect{y} }^d  \rho_r ( \vect{x} - \vect{y} ) 
	; \vect{x} \in \texttt{D}  \}
\end{align}
with weight $ m $-vector $ \vect{\alpha}^d 
= \sigi{r}  [ \Sigi{d} ]^{-1} ( \vect{r}^d - \mu_r \iv{m} ) 
=  [ \Sigi{d}^\rho ]^{-1} ( \vect{r}^d - \mu_r \iv{m} )
=( \alpha_1^d , \dots , \alpha_m^d )^T $.  Hence the predictor appears as a linear combination of the correlations between the spatial variable value to be predicted and the observation values $  \nu_i ( \vect{x} ) = \rho_r ( \vect{x} - \vect{x}_i^d ) ; i = 1, \dots, m $, called the observation kernel functions.
The corresponding prediction variance expression is,
\begin{align}
	\{  \sigi{p} ( \vect{x} ) &= \sigi{r}
	[ 1- \vect{\rho}_{xd} [ \Sigi{d}^\rho ]^{-1} \vect{ \rho}_{dx} ] \\  \nonumber
	&= \sigi{r} [  1 - \Sigma_{ \vect{y}' \in \texttt{M} } \Sigma_{ \vect{y}'' \in \texttt{M} }
	\beta_{ \vect{y}'  \vect{y}'' } \times
	\rho_r ( \vect{x} - \vect{y}' ) \rho_r ( \vect{x} - \vect{y}'' ) ]
	; \vect{x} \in \texttt{D}  \} 
\end{align}
with weights $ \beta_{ \vect{y}'  \vect{y}'' } = [ [ \Sigi{d}^\rho ]^{-1} ]_{i j} ; i,j = 1, \dots , m $ with $ (i,j) $ and $ ( \vect{y}' , \vect{y}'' ) $ being matching observation numbers and locations.
The $ m $-vector of weights in the Kernel predictor $ \vect{ \alpha}^d $ is dependent on the observation locations $ \texttt{M} $ and the actual observed values $ \vect{r}^d $, 
but independent of the prediction location $ \vect{x} \in \texttt{D} $. Hence these weights are dependent on the random observed values, and are therefore random. Note further that
$ \Expe{ \vect{ \alpha}^d } = 0 \iv{m} $ and 
$ \Vari{ \vect{ \alpha}^d } 
=  [ \sigi{r} ]^2  [ \Sigi{d} ]^{-1} =   \sigi{r}  [ \Sigi{d}^\rho ]^{-1} $.
The $ \vect{ \rho }_{xd} $-term only depend on the model parameters and hence is deterministic.

The weight $ m $-vector 
$ \vect{ \alpha }^d =  [ \Sigi{d}^\rho ]^{-1} ( \vect{r}^d - \mu_r \iv{m} ) $
is global and only has to be calculated once. The functional representation of the predictor appears as a weighted linear combination of the symmetric spatial correlation functions centred at the $ m $ observation locations. 
No spatial grid discretization is required.

\vspace{5 mm}
\noindent
\textbf{Summary }

To summarize - both spatial predictor expressions are exact and optimal predictors for the values of the spatial variable at the grid nodes - and hence identical. This is so for all valid Gaussian RF models, hence all non-negative definite spatial correlation functions. The functional representation of the Kernel predictor is defined for all $ \vect{x} \in \texttt{D} $ and  appears as the grid infill asymptotic limit, see \cite{Cressie2011},  for
the grid representation of the traditional Kriging predictor. This limit can be assessed by the Kernel predictor 
with computational demands dominated by the Cholesky decomposition of the observation inter-correlation 
$ ( m \times m ) $-matrix $ \Sigi{d}^\rho $. No finite grid representation is required.

Since the functional representation is the grid infill limit, it can be used to generate a grid representation in arbitrary design, for example by zooming into a specific area. Moreover,
the functional representation make it possible to calculate analytically exact and optimal predictions, with associated prediction variances, for any linear operator on the Gaussian RF.
Examples of such operators are integration over a specific area and differentiation in an arbitrary location, if it exists.

\vspace{5 mm}
\noindent
\textbf{Extensions }

It is worth noting that both predictor expressions can easily be extended to cover cases with observations being linear operators of the spatial variable with additive Gaussian observation errors. As example, consider a case with one locationwise observation with error, 
$ d_1 = r( \vect{x}_1^d ) + e_1 $,
one differential observation in the $ \vect{z} $-direction with error, 
$ d_2 = d / d \vect{u }_{ \vect{z} } \mbox{  } r( \vect{u} ) |_{ \vect{u} = \vect{x}_2^d } 
+ e_2 $, 
and one spatially integrated observation with error, 
$ d_3 = \int_{\texttt{D}_3} r( \vect{u} ) d \vect{u} + e_3 $ with 
$ \texttt{D}_3 \subset \texttt{D} $.
Assume that the error terms are centered Gaussian and independent with 
variances $ \sigi{1} $,  $ \sigi{2} $ and $ \sigi{3} $, respectively. The prediction-observation correlation function entries in the $m $-vector $ \vect{\rho}_{xd} $,
denoted the kernel functions, will then be
\begin{align*}
	[ \vect{\rho}_{xd} ]_1 = \nu_1 ( \vect{x} ) 
	&= \rho_r ( \vect{x} - \vect{ x}_1^d ) \\
	[ \vect{\rho}_{xd} ]_2 = \nu_2 ( \vect{x} ) 
	&=  \mbox{   } d / d \vect{u}_{ \vect{z} } \mbox{  } \rho_r ( \vect{x} - \vect{u} ) | _{ \vect{u} = \vect{x}_2^d }  \\
	[ \vect{\rho}_{xd} ]_3 = \nu_3 ( \vect{x} )
	&=  \int_{ \texttt{D}_3} \rho_r ( \vect{x} - \vect{u} ) d \vect{u} 
\end{align*}
The corresponding observation inter-correlation $ ( m \times m) $-matrix entries in 
$ \Sigi{d}^\rho $ are
\begin{align*}
	[ \Sigi{d}^\rho]_{1,1} 
	&= 1 + \frac{\sigi{1} }{ \sigi{r} } \\
	[ \Sigi{d}^\rho ]_{2,2}
	&=  \mbox{   } d^2 / d \vect{u}_{ \vect{z} } d \vect{v}_{ \vect{z} } \mbox{  }
	\rho_r ( \vect{u} - \vect{v} ) | _{ \vect{u} = \vect{v} = \vect{x}_2^d } 
	+ \frac{ \sigi{2}}{ \sigi{r} } \\
	[ \Sigi{d}^\rho ]_{3,3} 
	&=  \int_{ \texttt{D}_3 } \int_{ \texttt{D}_3 } 
	\rho_r ( \vect{u} - \vect{v} ) d \vect{v} d \vect{u} + \frac{ \sigi{3} }{ \sigi{r} } \\
	[ \Sigi{d}^\rho ]_{1,2} = [ \Sigi{d}^\rho ]_{2,1}
	&=   d / d \vect{u}_{ \vect{z} } \mbox{  }
	\rho_r (  \vect{x}_1^d - \vect{u} ) | _{ \vect{u} = \vect{x}_2^d } \\
	[ \Sigi{d}^\rho ]_{1,3} = [ \Sigi{d}^\rho]_{3,1} 
	&=   \int_{ \texttt{D}_3} \rho_r ( \vect{x}_1^d - \vect{u} ) d \vect{u}  \\
	[ \Sigi{d}^\rho ]_{2,3} = [ \Sigi{d}^\rho ]_{3,2}
	&=
	\int_{ \texttt{D}_3 }
	\mbox{  }   d / d \vect{u}_{ \vect{z} }
	\rho_r ( \vect{u} - \vect{v} ) 
	| _{ \vect{u} = \vect{x}_2^d }
	d \vect{v}
\end{align*}
Note that if all these terms can be calculated analytically, then the Kernel predictor will appear as a linear combination of the three observation kernel functions in
$ \vect{ \nu } ( \vect{x} ) = ( \nu_1 ( \vect{x} ) , \nu_2 ( \vect{x} ) , 
\nu_3  ( \vect{x} ) )^T $ which will provide an exact functional representation of the spatial variable under Gaussian assumptions. The corresponding prediction variance can also be represented on a functional form.
Model parameter inference is somewhat more complicated for this case with observation errors.

Return to the case with locationwise exact observations with predictor and predictor variance for the spatial variable in arbitrary location $ \vect{x}_+ \in \texttt{D} $ defined in Expression \ref{eq:GRF-Pred}. Assume that focus is not on locationwise prediction but rather
on either the derivative in direction $ \vect{z} $ of the spatial variable in an 
arbitrary location $ \vect{x}_+ \in \texttt{D} $ or
on the spatial average over an arbitrary volume $ \texttt{D}_p \in \texttt{D} $. 
Note that both these features are linear operators on the spatial variable, hence their predictors are identical to the corresponding operator on the locationwise predictor. For the Kernel predictor stored on a functional representation these predictors and prediction variances can be exactly defined, if they exist.
The derivative is predicted as,
\begin{align*}
	\hat{r}_{d_{ \vect{z}} }
	= d / d \vect{u}_{ \vect{z}} \mbox{  } \hat{r} ( \vect{u} ) |_{ \vect{u} = \vect{x}_+ } 
	= \Sigma_{ \vect{y} \in \texttt{M} } \mbox{ } \alpha_{ \vect{y} }^d \times
	d / d \vect{u}_{ \vect{z} }  \mbox{  } \rho_r ( \vect{u} - \vect{y} ) 
	|_{ \vect{u} = \vect{x}_+ }
\end{align*}
and the expression for the associated prediction variance is on differential quadratic form and can easily be developed. 
The spatial average is predicted as,
\begin{align*}
	\hat{ r }_{ \texttt{D}_p } = | \texttt{D}_p |^{-1} \int_{ \texttt{D}_p }
	\hat{ r } ( \vect{u} ) d \vect{u}  
	= \mu_r + | \texttt{D}_p |^{-1}
	\Sigma_{ \vect{y} \in \texttt{M} } \mbox{ } \alpha_{ \vect{y} }^d \times
	\int_{ \texttt{D}_p } 
	\rho_r ( \vect{u} - \vect{y} ) d \vect{u} 
\end{align*}
and the expression for the associated prediction variance is on integrated quadratic form and can easily be developed. 
Note that these expressions can be calculated analytically if the integral and derivative of the spatial correlation functions can be assessed analytically.

For the traditional Kriging predictor stored on a grid representation, the corresponding predictors can only be defined numerically as 
the slope in the grid in $ \vect{z } $-direction from 
location $ \vect{x}_+ $ and
the average of the locationwise predictions in grid nodes inside 
$ \texttt{D}_p $. 
In order to assess the corresponding prediction variances, not only predictions and variances in the grid nodes involved must have been stored, but also all the kriging weights used in these grid nodes. The Kernel predictor appears as clearly superior to the traditional Kriging predictor for these cases where the objective is to predict linear operators on the spatial variable.

\vspace{5 mm}
\noindent
\newpage
\textbf{Example A}

\begin{figure} 
	\centering
	\begin{subfigure}[ht]{0.4\textwidth}
		\includegraphics[width=1.0\textwidth]{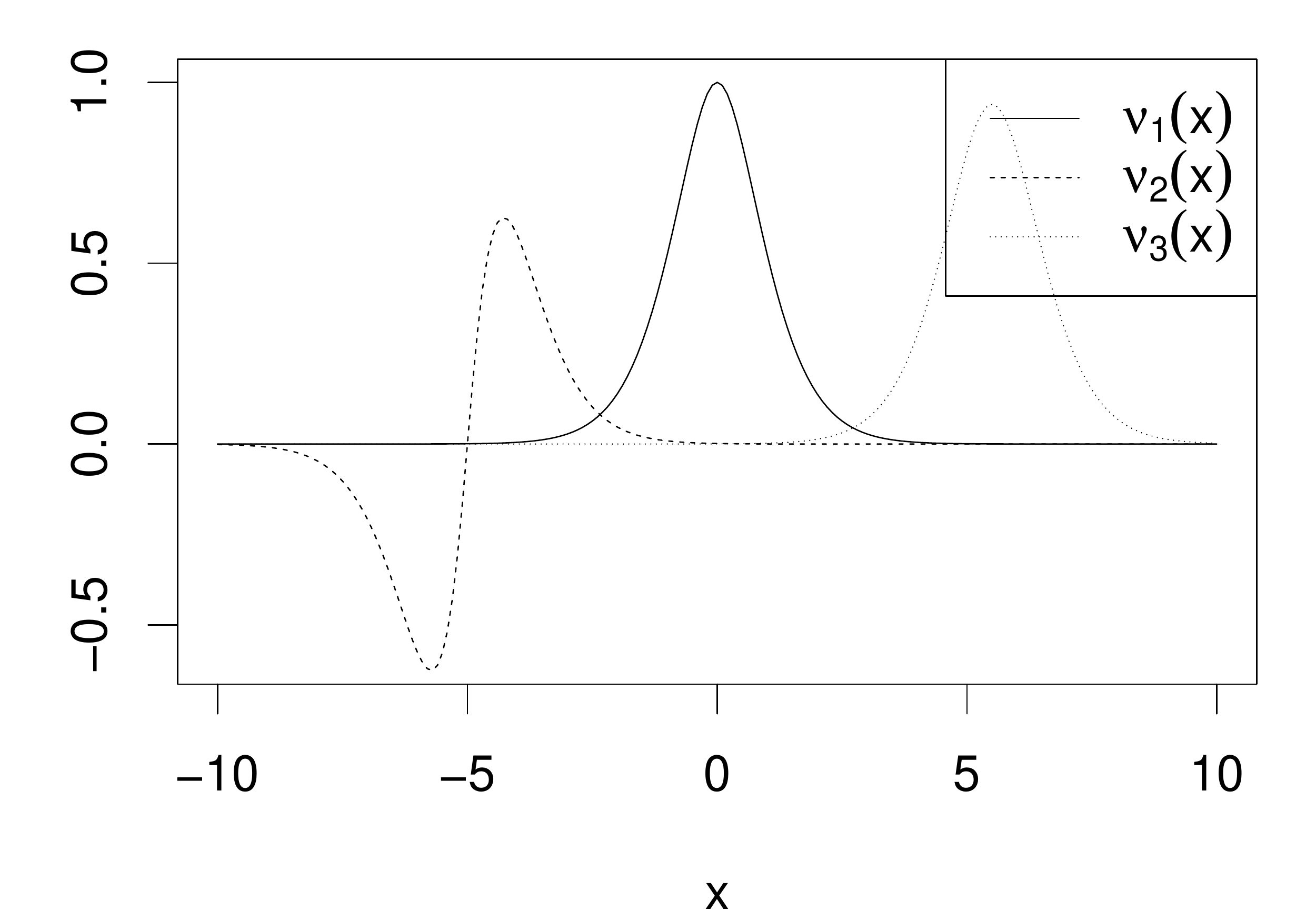}
		\caption{Range $ 1.0 $.} 
	\end{subfigure}
	\begin{subfigure}[ht]{0.4\textwidth}  
		\includegraphics[width=1.0\textwidth]{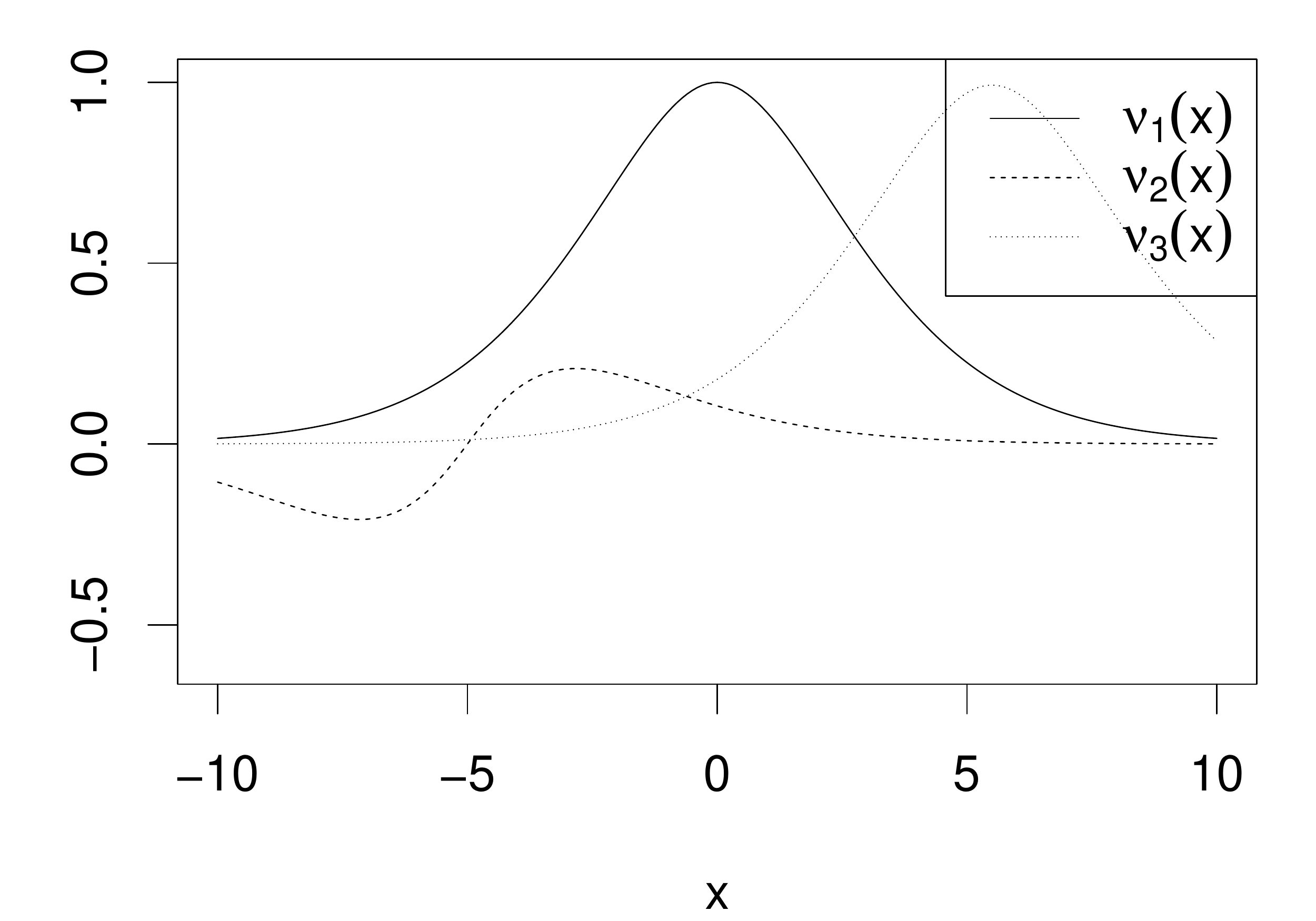}
		\caption{Range $ 3.0 $. } 
	\end{subfigure}
	\caption{Observation kernel functions: Gaussian RF model with Matern spatial correlation function - shape $ 5/2 $. } 
	\label{fig:Pred-Obs}
\end{figure}

\begin{figure}  
	\centering
	\begin{subfigure}[ht]{0.4\textwidth}
		\includegraphics[width=1.0\textwidth]{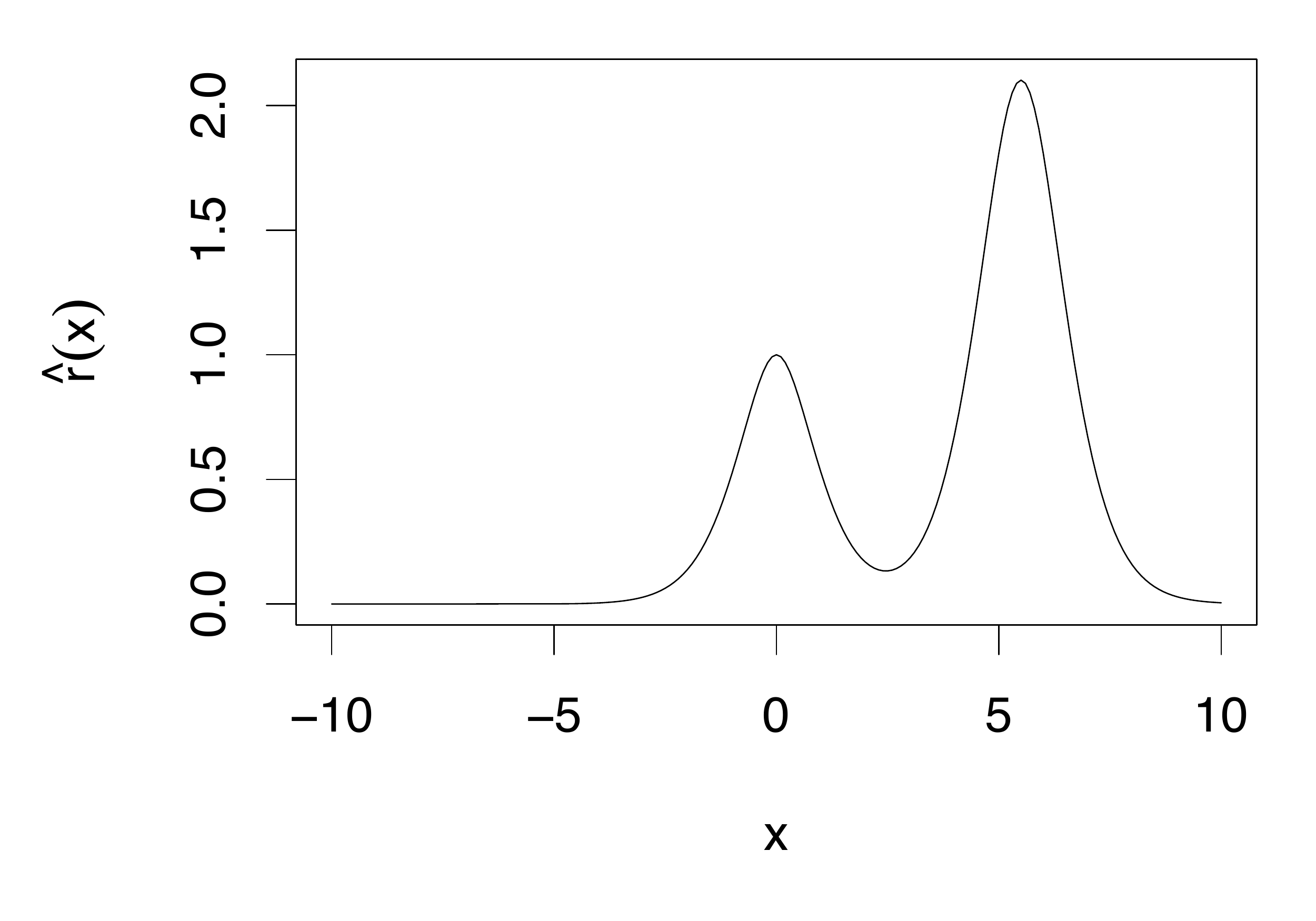}
		\caption{Kernel prediction}
	\end{subfigure}
	\begin{subfigure}[ht]{0.4\textwidth}
		\includegraphics[width=1.0\textwidth]{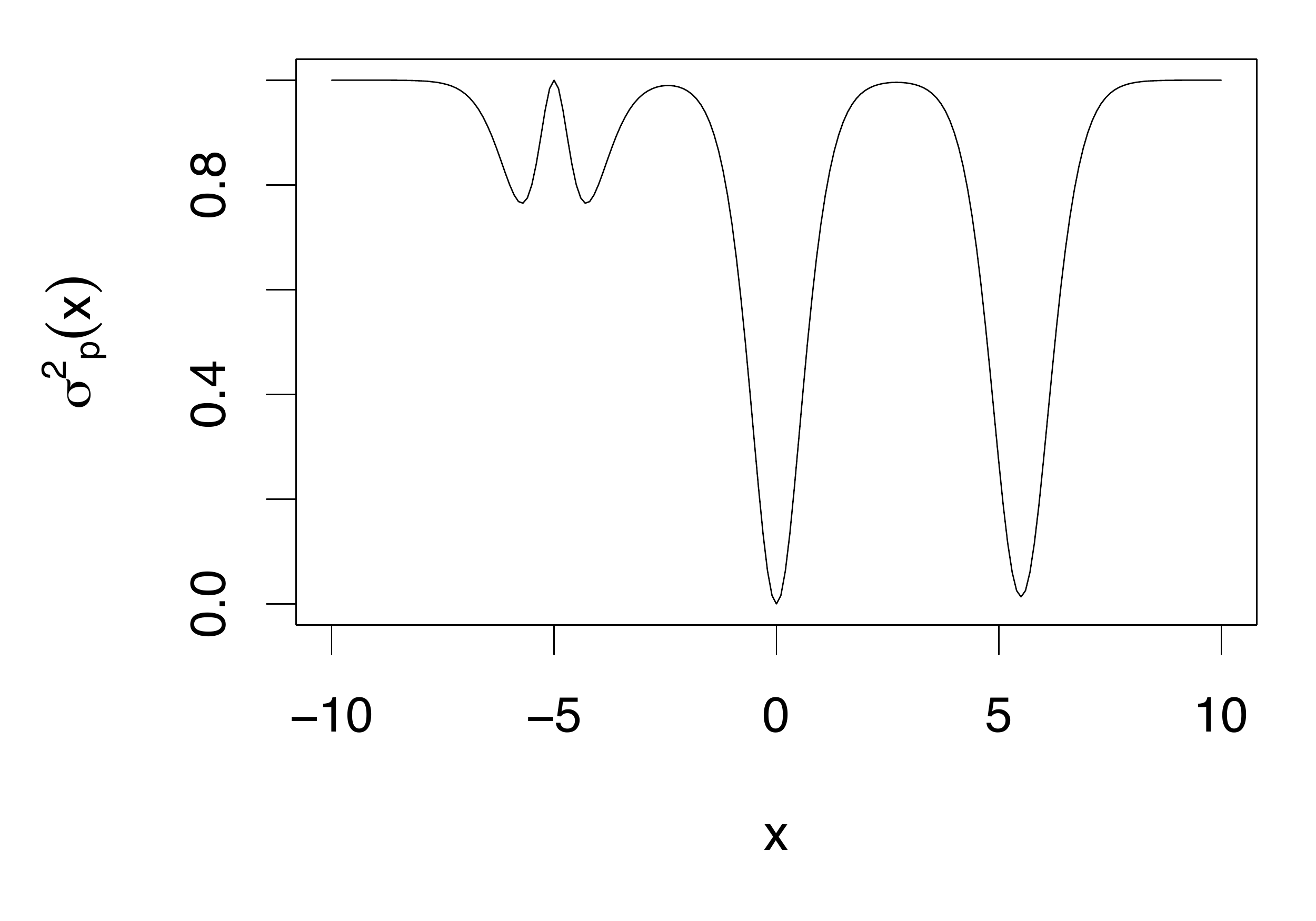}
		\caption{Kernel prediction variance}
	\end{subfigure}
	\caption{Kernel predictor: Gaussian RF model with Matern spatial correlation function - 
		shape $ 5/2 $ and range $ 1.0 $ - observations $ ( 1,0,2) $. } 
	\label{fig:Func-Pred-1.0-A}
\end{figure}

\begin{figure}  
	\centering
	\begin{subfigure}[ht]{0.4\textwidth}
		\includegraphics[width=1.0\textwidth]{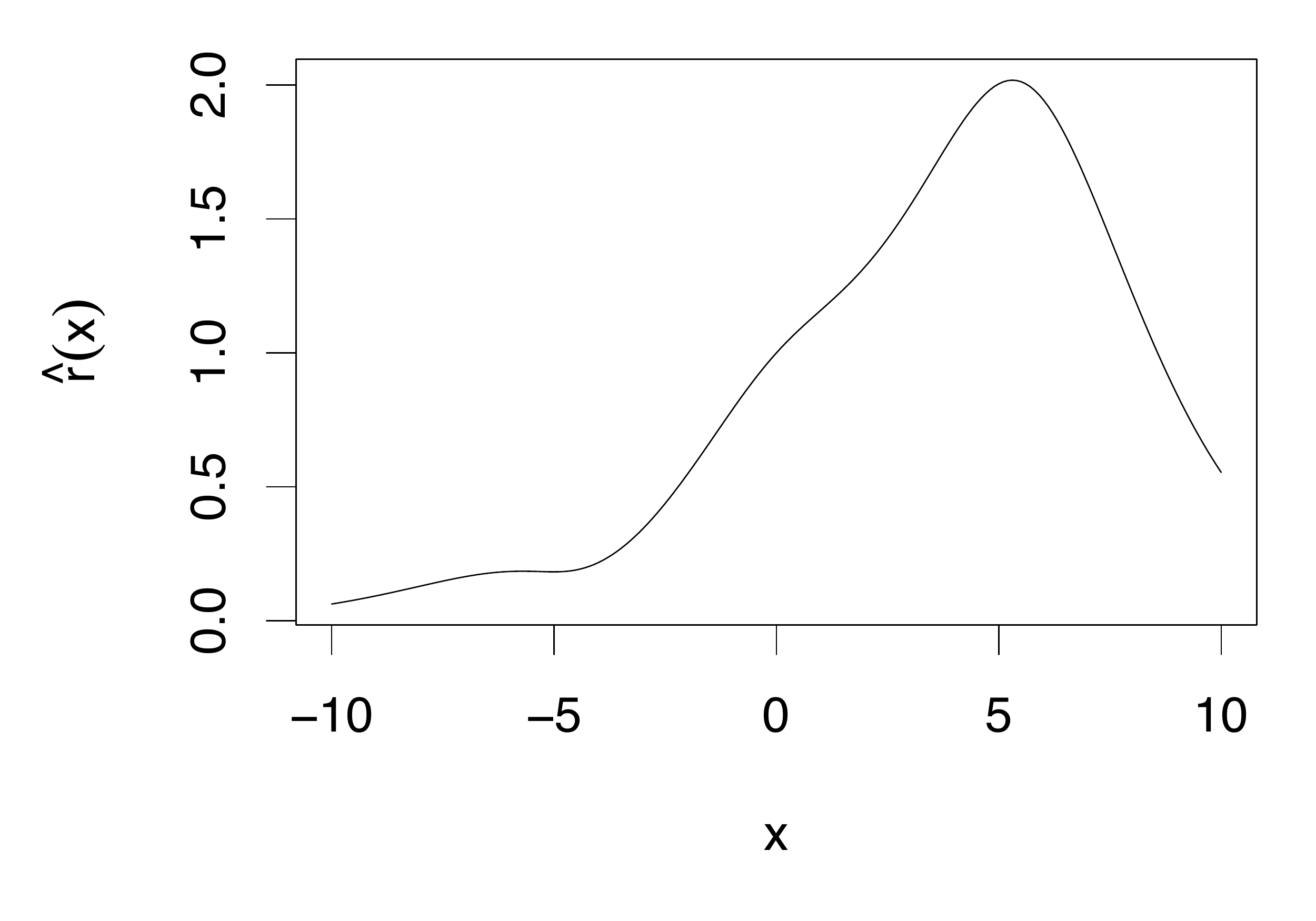}
		\caption{Kernel prediction}
	\end{subfigure}
	\begin{subfigure}[ht]{0.4\textwidth}
		\includegraphics[width=1.0\textwidth]{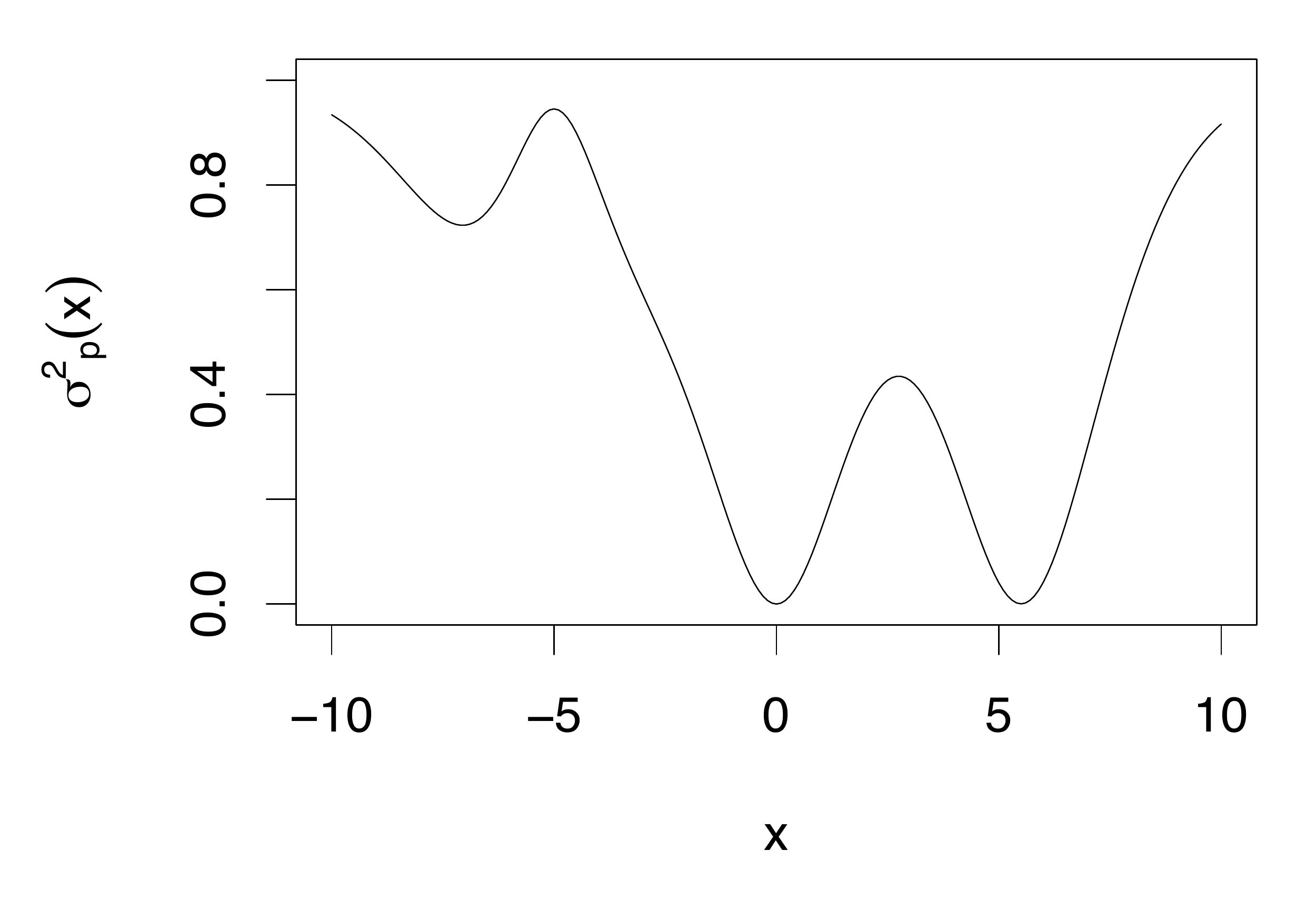}
		\caption{Kernel prediction variance}
	\end{subfigure}
	\caption{Kernel predictor: Gaussian RF model with Matern spatial correlation function - 
		shape $ 5/2 $ and range $ 3.0 $ - observations $ ( 1,0,2) $. } 
	\label{fig:Func-Pred-3.0-A}
\end{figure}

The example displays the flexibility of the Kernel predictor in a case with a spatial variable defined with a one-dimensional reference domain 
$ \{ r ( x ) ; x \in [-10, 10 ] \subset \mathbb{R} \} $. The stationary Gaussian RF model has expectation level $ \mu_r = 0.0 $ and variance level $ \sigi{r} = 1.0 $, with spatial correlation function belonging to the Matern class of functions with shape parameter 
$  5/2 $ and range parameter $ \tau_M \in \mathbb{R}_+ $, hence the spatial correlation function is
\begin{align*}
	\rho_r ( \tau ) = [ 1 + 5^{1/2} \tau + 5/3 \tau^2 ] \exp \{ - 5 ^{ 1/2 } \tau \}  
	; \tau = | x' - x'' | / \tau_M .
\end{align*}
The observation set  represented by the $ 3 $-vector $ \vect{d} $ on which the spatial prediction must be based consists of one exact locationwise observation at $ x=0 $, one exact derivative  observation at $ x = -5 $, and one exact average observation over the interval $ x \in [5, 6 ] $,
\begin{align*}
	d_1 &= r ( 0) \\
	d_2 &= d / dx \mbox{  }r (x ) |_{x= -5} \\
	d_3 &= \int_5^6 r (u ) du. 
\end{align*}
In order to define the Kernel predictor one needs to calculate the observation kernel functions $ \vect{ \nu } ( x)  = ( \nu_1 (x) , \nu_2 ( x) , \nu_3 (x ) )^T   $ and the entries in the observation inter-correlation $ ( 3 \times 3 ) $-matrix $ \Sigi{d}^\rho $. Expressions for these terms are calculated, but we only present the expressions for the observation kernel functions parametrized by $ \kappa_0 = \sqrt{5} / \tau_M $,
\begin{align*}
	\nu_1 ( x) &= \rho_r ( | x -0 |/ \tau_M )  \\
	&= 
	\left\{ 
	\begin{array}{ll}
		.[ 1 + \kappa_0 x + \kappa_0^2  x^2 / 3] \exp \{ - \kappa_0 x \}  &  ; 0 \leq  x \leq 10 \\
		.[1 - \kappa_0 x +  \kappa_0^2  x^2 / 3] \exp \{ \kappa_0 x \}  & ; -10 \leq x < 0
	\end{array}
	\right.                            \\
	&=
	[ 1 + \kappa_0 |x| + \kappa_0^2  |x|^2 / 3 ] \exp \{ - \kappa_0 |x| \} \mbox{   }  ; -10 \leq x \geq 10 \\
	\nu_2 ( x ) &= d / du \mbox{  } \rho_r ( | x - u | / \tau_M  ) | _{ u = -5 }    \\
	&=
	\left\{
	\begin{array}{ll}
		\kappa_0^2 /3 [ (x+5) + \kappa_0  (x+5)^2 ] \exp \{ - \kappa_0 (x+5) \} 
		& ; -5 \leq x \leq 10 \\
		\kappa_0^2 /3 [ (x+5) - \kappa_0 (x+5)^2 ] \exp \{ \kappa_0 (x+5) \} 
		&  ; -10 \leq x < -5 
	\end{array}
	\right.  \\
	&=
	\text{sign} \{ x+5 \} \times \kappa_0^2 /3 [ |x+5| + \kappa_0 |x+5|^2 ] 
	\exp \{ - \kappa_0 |x+5| \} 
	\mbox{   } ; -10 \leq x \leq 10 \\
	\end{align*}
	\begin{align*}
	\nu_3 (x ) &= \int_5^6 \rho_r ( | x-u | / \tau_M ) du \\
	&=
	\left\{ 
	\begin{array}{ll}
		- \kappa_0 /3 \times [ ( 8/\kappa_0^2 + 5   (x-5) / \kappa_0 + (x-5)^2 ) 
		\exp \{ - \kappa_0  (x-5) \} & \\
		\qquad - ( 8/ \kappa_0^2 + 5 (x-6) / \kappa_0 + (x-6)^2) \exp \{ - \kappa_0 (x-6) \} ] 
		& ; 6 \leq x \leq 10  \\ 
		- \kappa_0 /3 \times  [ ( 8/ \kappa_0^2 + 5 (x-5) / \kappa_0+ (x-5)^2 ) 
		\exp \{ - \kappa_0 (x-5) \} - 8/ \kappa_0^2 ] & \\
		\qquad- \kappa_0 /3 \times [ ( 8/ \kappa_0^2 + 5 (6-x) / \kappa_0 + (6-x)^2 ) 
		\exp \{ - \kappa_0 (6-x)  \} - 8/ \kappa_0^2 ] 
		& ; 5 < x < 6 \\ 
		- \kappa_0  / 3 \times [ ( 8/ \kappa_0^2 + 5 (6-x) / \kappa_0 + (6-x)^2 ) 
		\exp \{ - \kappa_0 (6-x) \}& \\
		\qquad- ( 8/ \kappa_0^2 + 5 (5-x) / \kappa_0 + (5-x)^2 ) \exp \{ - \kappa_0 (5-x) \}]  
		& ; -10 \leq x \leq 5
	\end{array}
	\right.  \\
	&=
	\left\{ 
	\begin{array}{ll}
		\kappa_0 /3 \times [ ( 8/ \kappa_0^2 + 5 |x-6| / \kappa_0 + |x-6|^2 ) 
		\exp \{ - \kappa_0 |x-6| \} &\\
		\qquad- ( 8/ \kappa_0^2 + 5 |x-5| / \kappa_0+ |x-5|^2) \exp \{ - \kappa_0 |x-5| \} ] 
		& ; 6 \leq x \leq 10  \\ 
		\kappa_0 /3 \times [ 16 / \kappa_0^2 & \\
		\qquad- ( 8/ \kappa_0^2 + 5 (x-5) / \kappa_0 + (x-5)^2 ) 
		\exp \{ - \kappa_0 (x-5) \}  & \\
		\qquad- ( 8/ \kappa_0^2 + 5 (6-x) / \kappa_0+ (6-x)^2 ) 
		\exp \{ - \kappa_0 (6-x)  \} ] 
		& ; 5 < x < 6 \\ 
		\kappa_0 /3 \times [ ( 8/ \kappa_0^2 + 5 |x-5| / \kappa_0+ |x-5|^2 ) 
		\exp \{ - \kappa_0 |x-5| \} &\\
		\qquad- ( 8/ \kappa_0^2 + 5 |x-6| / \kappa_0 + |x-6|^2) 
		\exp \{ - \kappa_0 |x-6| \} ]  
		& ; -10 \leq x \leq 5
	\end{array}
	\right. 
\end{align*}
These functions are analytically assessed and the spatial Kernel predictor will appear as a linear combination of these functions. The actual shape of the three observation kernel functions 
$ \vect{ \nu } ( x ) = ( \nu_1 ( x ) , \nu_2 ( x ) , \nu_3 ( x ) )^T $
for the range parameter $ \tau_M = 1.0 $ are displayed in Figure \ref{fig:Pred-Obs}.a. 
Note that $ \nu_1 ( x ) $, centred at $ x = 0 $, indicates that the value 
at $ x= 0 $, $ r ( 0 ) $,
and the observation $ d_1 $ have correlation $ 1.0 $. The function, $ \nu_2 ( x ) $,
centred at $ x = -5 $, indicates that the value at $ x = -5 $, $ r ( -5 ) $, and the observation $ d_2 $ are uncorrelated. Recall that in a given location the Gaussian RF variable and its derivative are independent. The correlation does increase/decrease to the right/left 
of $ x = -5 $. The function, $ \nu_3 ( x ) $, is centred at $ x= 5.5 $ being the center location of the averaging interval. The corresponding observation $ d_3 $ is positively correlated with the Gaussian RF values, but nowhere is the correlation equal $ 1.0 $ caused by the averaging over an interval.
Let the observed values be $ \vect{d} = ( d_1 , d_2 , d_3 )^T = ( 1, 0 ,2 )^T $.
The weights in the predictor are calculated 
as $ \vect{ \alpha }^d = [ \Sigi{d}^\rho ]^{-1} \vect{d} $, and the actual values are computed to be $ \vect{ \alpha }^d = (0.9992,-0.00085, 2.23876)^T $. The Kernel predictor is then,
\begin{align*}
	\{ \hat{r} ( x ) = [\vect{ \nu } ( x ) ]^T \vect{ \alpha }^d 
	= \Sigma_{i=1}^3  \alpha_i^d \nu_i (x ) ; x \in [ -10 , 10 ] \},
\end{align*}
with associated prediction variance,
\begin{align*}
	\{ \sigi{p} (x ) = 1 - [ \vect{ \nu } (x) ]^T [ \Sigi{d}^\rho ]^{-1} \vect{ \nu } ( x ) 
	= 1 - \Sigma_{i=1}^3 \Sigma_{ j=1 }^3 \beta_{ij} \nu_i (x ) \nu_j (x ) 
	; x \in [ -10, 10 ] \},
\end{align*}
with $ \beta_{ij} = [ [ \Sigi{d}^\rho ]^{-1} ]_{ij} ; i,j = 1, \dots , 3 $.
In Figure \ref{fig:Func-Pred-1.0-A} the results from the Kernel predictor is displayed. The left display contains the spatially continuous  prediction with the exact observations correctly reproduced.
The predictions approach the expectation level $ \mu_r = 0 $ between the observations since the range of the spatial correlation function is short, hence the information in the observations do not extend far.
The right display contains the spatially continuous prediction variances. The exact observation of the spatial variable at $ x = 0 $ entails that the prediction variance is zero, while the other two observations do not exactly define the spatial variable itself, in spite the observations are made without error, hence the prediction variance is larger than zero.
The parameters of the Gaussian RF model is adjusted such that the correlation range is longer, the range parameter is set to $ \tau_M = 3.0 $. The corresponding observation kernel functions $ \vect{ \nu } ( x ) $ are displayed in Figure 
\ref{fig:Pred-Obs}.b.  The actual observations are kept unchanged, 
$ \vect{d} = ( 1, 0 , 2 )^T $. Also the weights 
$ \vect{ \alpha }^d $ are influenced by the range parameter, and the actual weight values are computed to be $ \vect{ \alpha }^d = (0.7092, -0.4748, 1.9045)^T $. The results from the Kernel predictor are displayed in Figure \ref{fig:Func-Pred-3.0-A}. The prediction in the left display reproduces the observations, and it is smoother than the prediction based on the short-range model. The prediction variances in the right display also appear as smoother than the short-range results.
\begin{figure}  
	\centering
	\begin{subfigure}[ht]{0.4\textwidth}
		\includegraphics[width=1.0\textwidth]{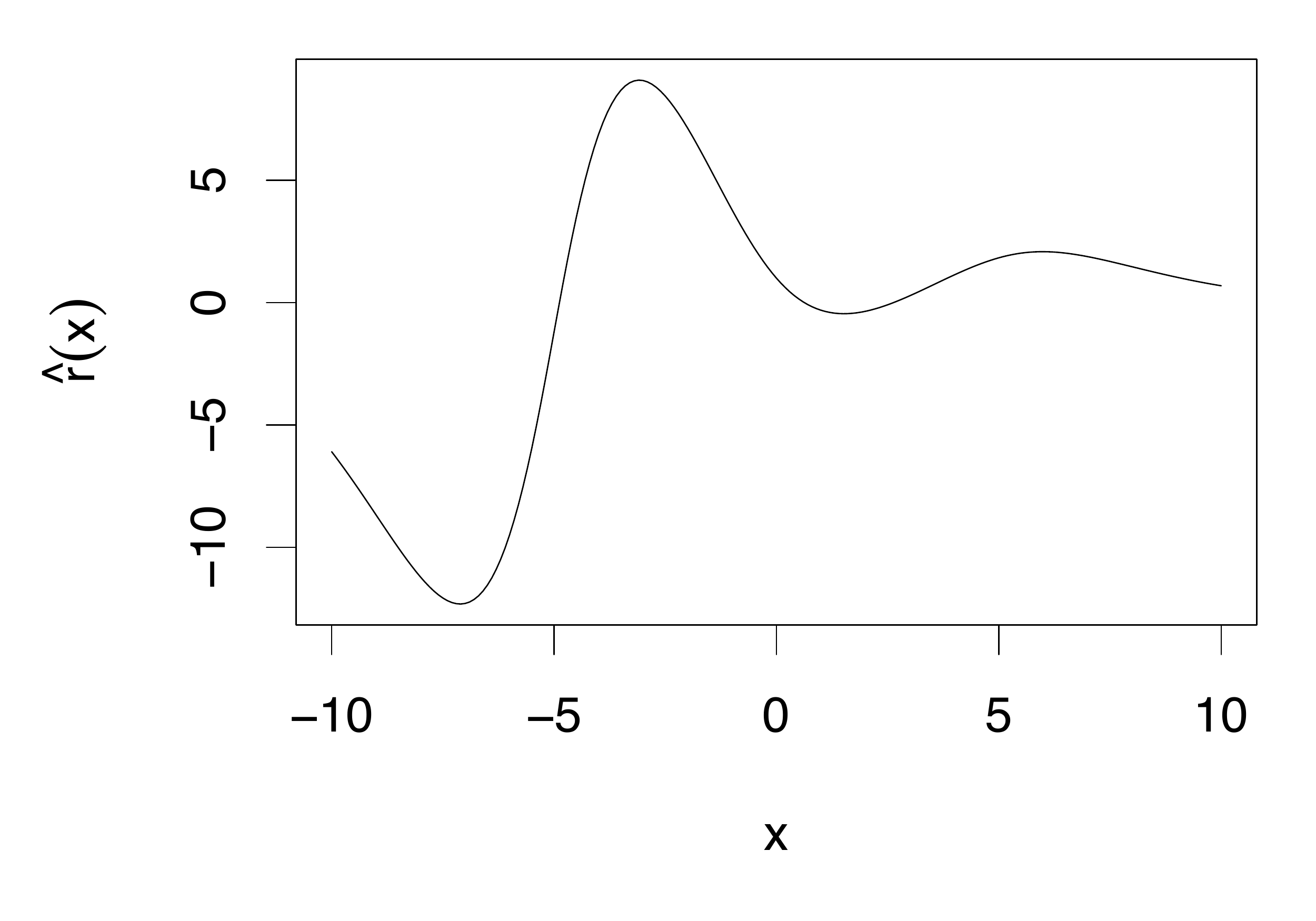}
		\caption{Kernel prediction}
	\end{subfigure}
	\begin{subfigure}[ht]{0.4\textwidth}
		\includegraphics[width=1.0\textwidth]{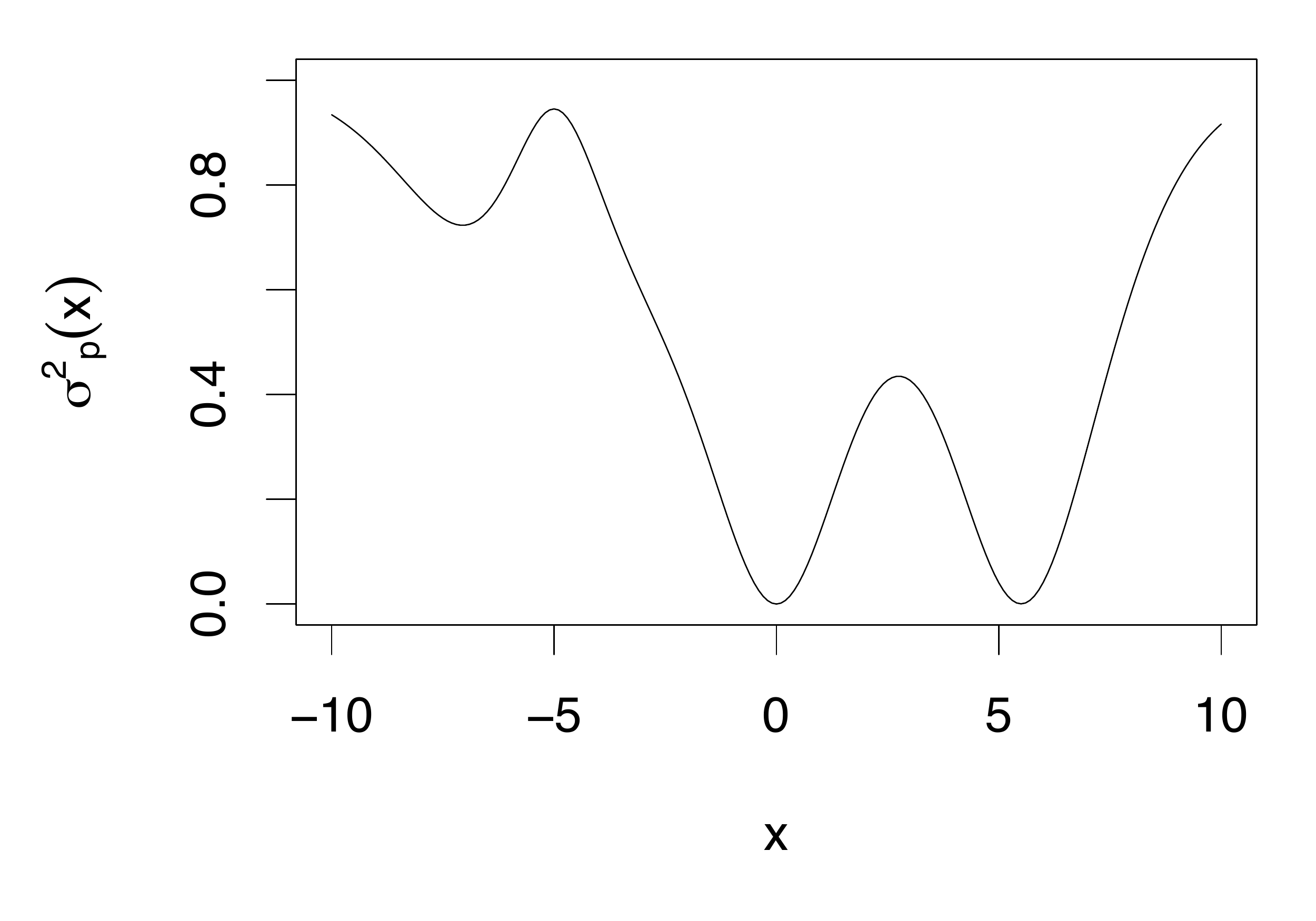}
		\caption{Kernel prediction variance}
	\end{subfigure}
	\caption{Kernel predictor: Gaussian RF model with Matern spatial correlation function - 
		shape $ 5/2 $ and range $ 3.0 $ - observations $ ( 1,10,2) $. } 
	\label{fig:Func-Pred-3.0-B}
\end{figure}
Lastly, the same Gaussian RF model parameters as listed above is used, see Figure 
\ref{fig:Pred-Obs}.b. The observation values are changed however, to 
$ \vect{d} = ( 1, 10, 2 )^T $ which entails that the gradient of the spatial variable at 
$ x = -5 $ is observed to be much larger. This change 
does influence the weights which are calculated 
to be $ \vect{ \alpha }^d = ( -5.4759, 57.0169, 2.6231)^T $.
In Figure \ref{fig:Func-Pred-3.0-B} the results from the Kernel predictor are presented. The left display contains the prediction which reproduces the observed values, and the change of gradient at $ x = -5 $ does dramatically change the prediction compared to the prediction with gradient zero. The prediction variances in the right display is identical to the one for gradient zero, however, since for Gaussian RF models the prediction variances are independent of the actual observed values.

\section{Large observation sets }

Both predictors discussed in Section \ref{sec:Spat-Pred} require that 
the observation inter-correlation $ ( m \times m) $-matrix 
$ \Sigi{d}^\rho $ is inverted, which is usually made by Cholesky decomposition. For large observation sets, hence large $ m $, this decomposition may be computationally demanding since in general matrix inversion requires computer processing of order $ m^3 $. If the 
$ ( m \times m)$-matrix is sparse, however, efficient algorithms for Cholesky decomposition of orders $ m^{ 3/2} $ for reference domain $ \texttt{D} \subset \mathbb{R}^2 $ and
$ m^2 $ for $ \texttt{D} \subset \mathbb{R}^3 $
are available, see \cite{GMRFbook}. For huge observation sets, hence huge $ m $, Cholesky decomposition of the $ ( m \times m ) $-matrix, even after sparsification, may be computationally unfeasible. Then approximations must be used.

\vspace{5 mm}
\noindent
\textbf{The Traditional Kriging Predictor }

The traditional Kriging predictor is frequently used in applications with large observation sets where the Cholesky decomposition of the $ (m \times m ) $-matrix $ \Sigi{d}^\rho $ is challenging. The numerical mitigation trick used is to replace the matrix by a sparse approximation,
\begin{align*}
	\matr{\Sigma}_d^{\rho \ast} = \matr{T} \otimes \Sigi{d}^\rho
\end{align*}
where the tapering $ (m \times m ) $-matrix $ \matr{T} $ is a user-specified non-negative definite sparse matrix, and $ \otimes $ is elementwise multiplication, see \cite{Stein2013}. 
Then the approximation
$ \matr{\Sigma}_d^{\rho \ast } $ will be non-negative  definite and sparse, hence Cholesky decomposition can be made more efficiently.

The Gauss Markov RF model in \cite{GMRFbook}, defined on a spatial grid, is also constructed to be able to handle large sets of observations. The spatial effects in the formulation is represented by a non-negative definite 
precision $ ( n \times n) $-matix $ \matr{\Omega}_r $ which is defined as the inverse of the 
covariance $ ( n \times n ) $-matrix of the entire grid representation, 
hence  $ \matr{\Omega}_r = [  \Sigi{r} ]^{-1} $.  The Gauss Markov effect is obtained by assuming a certain sparsity in the precision $ ( n \times n ) $-matrix $ \matr{\Omega}_r $. If the observations are assumed to be locationwise, exact and located in a sub-set of the grid nodes, the corresponding 
observation inter-precision $ ( m \times m ) $-matrix $ \matr{\Omega}_d $ will be a sub-matrix of 
$ \matr{\Omega}_r $ and also be non-negative definite and sparse.
Spatial prediction requires that the inverse of the sparse $ ( (n-m) \times (n-m)) $-matrix
$ \matr{\Omega}_{r \backslash d} $, containing covariances between unobserved values in the grid, is assessed and Cholesky decomposition can be made efficiently.

The Gauss Markov RF model is defined on a specific spatial grid. The model is actually a lattice model, not a RF model.
The formulation is designed for cases with exact observations located in a sub-set of the grid nodes. If observations are linear combinations of values in grid nodes and/or have observation errors, spatial prediction requires the full sparse precision $ ( n \times n) $-matrix 
$ \matr{\Omega}_r  $ to be Cholesky decomposed. 
Recall that it is reasonable to have $ n \gg m $. 
Conditioning on observations located off grid cannot be made since the model is defined on the grid only. 
This lack of observation generalization can be crucial in many applications.

In studies with huge observation sets, hence huge $ m $, 
Cholesky decomposition of the $ ( m \times m) $-matrix $ \Sigi{d}^\rho $ can be computationally unfeasible  even after sparsification by tapering. In traditional geostatistics the mitigation strategy in these cases is usually to enforce localization, by assuming
some sort of Markov property, see \cite{Journel1978}, \cite{Datta2016} 
and \cite{Asfaw2016},
\begin{align*}
	\pdfc{ r( \vect{x}_+ ) }{\vect{ r }^d } 
	= \pdfc{ r ( \vect{x}_+ ) }{ r_{\vect{y} }^d ; \vect{y} \in \texttt{M}_+^\Delta }
\end{align*}
with the neighborhood observation location sub-set 
$ \texttt{M}_+^\Delta : \{ \vect{y} | \vect{y} \in \texttt{M} ; 
| \vect{y} - \vect{x}_+ | \leq \Delta  \} $, of size $ m_+^\Delta $. This sub-set contains the observation locations closer to $ \vect{x}_+ $ than $ \Delta $.  
The neighborhood sub-set can alternatively be defined as the 
$ m_+ $ closest observations or as a direction-stratified sub-set.
The Vecchia concept discussed in \cite{Katzfuss2021} can be used to define the actual neighborhoods used in the localized Kriging predictor. 

The localized traditional Kriging predictor discretized to the grid $ \texttt{L} $ under these Markovian assumptions, is
\begin{align*}
	\{  \hat{  r}^\ast ( \vect{x} ) = \mu_r 
	+ \Sigma_{ \vect{y} \in \texttt{M}_x^\Delta } 
	\alpha_{\vect{y} }^{x \ast} (r_{\vect{y} }^d - \mu_r ) 
	; \vect{x} \in \texttt{L} \subset \texttt{D} \}
\end{align*}
with associated prediction variances,
\begin{align*}
	\{ \sigma_p^{ 2 \ast}  ( \vect{x} ) = \sigi{r} 
	[ 1 - \vect{\alpha }^{x \ast T}  \Sigi{d_x^\Delta}^\rho    \vect{\alpha }^{x \ast} ] 
	; \vect{x} \in \texttt{L} \subset \texttt{D}  \}
\end{align*}
where the weight $ m_x^\Delta $-vector
$  \vect{ \alpha}^{x \ast } 
= [ \Sigi{d_x^\Delta}^\rho ]^{-1} \vect{ \rho }_{d_x^\Delta x }$
with $ \Sigi{d_x^\Delta}^\rho $ being the observation inter-correlation
$ ( m_x^\Delta \times m_x^\Delta ) $-matrix of the observations in the location sub-set 
$ \texttt{M}_{ \vect{x} }^\Delta $. The correlation $ m_x^\Delta $-vector 
$ \vect{ \rho }_{d_x^\Delta x } $ is defined similarly as correlations between  the observations in the location sub-set 
$ \texttt{M}_{ \vect{x} }^\Delta $ and the value to be predicted $ r( \vect{x} ) $. This predictor is of course exact and optimal for
$ r ( \vect{x} ) $ at arbitrary location $ \vect{x} \in \texttt{D} $ given the observations in the observation location sub-set 
$ \texttt{M}_{ \vect{x} }^\Delta $, but since parts of the global observation location set $ \texttt{M} $ 
is ignored, the predictor and prediction variances are not optimal in the global sense.
In real studies also the expectation and variance levels $ ( \mu_r , \sigi{r}  ) $ must be estimated, and this inference is usually also made locally by using 
Expression \ref{eq:MLE-ExpVar}.

Note that for global traditional Kriging predictors the observation inter-correlation
$ ( m \times m ) $-matrix $ \Sigi{d}^\rho $ only needs to be Cholesky decomposed once, since it is common for all weight $ m $-vectors in each grid node. Whenever localized traditional Kriging predictors are used, the localized observation inter-correlation 
$ ( m_x^\Delta \times m_x^\Delta ) $-matrices $ \Sigi{d_x^\Delta}^\rho $ 
may be unique for each grid node $ \vect{x} \in \texttt{L} $. Each of  these matrices need to be inverted for the corresponding grid node, hence $ n $ matrix decompositions.
Moreover, for huge observation sets, $ m $ huge, the grid must be relatively dense, hence $ n \gg m $ will be very huge.

The computer demands can be challenging, both processing and storage. The processing time is proportional to the grid size $ n $, but the algorithm is suitable for parallell processing.
Both prediction and prediction variance need to be stored in each grid node. 
It is particularly challenging if the spatial reference domain is three-dimensional, and in spatio-temporal studies.

The localized Kriging predictor is exact in the sense that if an exact observation is available in a grid node, the observed value will be exactly predicted. For observations located off-grid the values will not be exactly reproduced. The spatial characteristics of the localized predictor will be distorted since artificial discontinuities will occur whenever extreme-valued observations are included/excluded of the neighborhood location sub-set. It is difficult to quantify this approximation distortion.

\vspace{5 mm}
\noindent
\textbf{The Kernel Predictor }

The Kernel predictor also requires the observation inter-correlation $ ( m \times m ) $-matrix $ \Sigi{d}^\rho $ to be inverted, usually through Cholesky decomposition. For large observation sets, with $ m $ large, this inversion is computationally demanding. The recommended mitigation strategy for this case is to constrain the model space for the spatial variable $ \{ r ( \vect{x} ) ; \vect{x} \in \texttt{D}  \} $.
The previous discussion of the Kernel predictor only requires a valid stationary Gaussian RF model, hence all non-negative definite spatial correlation functions are eligible. In order to obtain a sparse observation inter-correlation $ ( m \times m ) $-matrix the Gaussian RF is assumed to have a finite-range non-negative definite spatial correlation function, hence
\begin{align*}
	\rho_r ( \vect{\tau} ) = 
	\left\{
	\begin{array}{lll}
		1.0   &  | \vect{ \tau } | = 0.0  \\
		( -1.0 , 1.0 ) &  0.0 < | \vect{ \tau } | < \tau_0 \\
		0.0   &  | \vect{ \tau } | \geq \tau_0
	\end{array}
	\right.
\end{align*}
with finite-range value $ \tau_0 < \infty $. This class of finite-range correlation functions is large, see \cite{Gneiting2002} and a short discussion in Section \ref{sec:Corr-Func}.
A suitable choice of range value $ \tau_0 $ will cause the observation inter-correlation $ (m \times m ) $-matrix $ \Sigi{d}^\rho $ to be sparse, and sometimes even blocked. Efficient algorithms can then be used for Cholesky decomposition and hence the inverse of 
$  \Sigi{d}^\rho $ can be calculated.

Moreover, the spatial Kernel predictor, see Expression \ref{eq:KP-Pred}, will for the finite-range Gaussian RF model be a local predictor since 
$ \rho_r ( \vect{ \tau } ) = 0 $ for $ | \vect{ \tau } | > \tau_0 $, hence
\begin{align*}
	\{ \hat{r} ( \vect{x} ) = \mu_r +
	\Sigma_{ \vect{y} \in \texttt{M}_x^{\tau_0 } }
	\alpha_{ \vect{y} }^d  \rho_r ( \vect{x} - \vect{y} ) 
	; \vect{x} \in \texttt{D} \}
\end{align*}
with the observation location 
sub-set $ \texttt{M}_x^{\tau_0 } 
= \{ \vect{y} | \vect{y} \in \texttt{M} , | \vect{x} - \vect{y}| < \tau_0 \} $ containing observation locations which are closer than $ \tau_0 $ to location $ \vect{x} $. This spatial, local Kernel predictor is exact and optimal for the specified finite-range Gaussian RF model.

Several basis function models are used to represent RF continuously in space.
In \cite{Cressie2008} and \cite{Cressie2022} a finite support basis function model is discussed, but the model needs to be represented on a underlying spatially discretized grid which 
must be relatively dense to represent the observations reliably and to avoid unacceptable non-stationarity in the corresponding prior model.
The conditioning requires Cholesky decomposition of a matrix defined by the dimensionality of the observations $ m $, which can be specified to be sparse.
The models are in practice on a grid representation and the posterior Gaussian coefficient model  in the dimension of the grid $ n $, must be assessed.
These computational requirements make them less attractive than the Kernel predictor defined above.
Refinement of this finite support basis function model is made in \cite{Nychka2015} by using a multi-level grid, which allows more computationally efficient conditioning.

The Stochastic Partial Differential Equation (SPDE) model for a RF, see \cite{Lindgren2011}
and \cite{Lindgren2022}, is also defined continuously in space, and it is demonstrated that it exhibits a first-order Matern spatial correlation structure. Moreover, it is claimed that the continuous representation has Markovian characteristics. The formulation is however such that a basis function grid representation, usually a triangular planar one, on a Gauss Markov form is used for spatial prediction. Several numerical approximations are required to make efficient spatial predictions based on the SPDE model. The fact that it is only defined for one specific spatial model, the many approximations used in prediction, and the need to involve matrices defined by the full grid  dimension $ n $, make the SPDE predictor less attractive than the Kernel predictor.

Return to the Kernel predictor and consider studies with huge observation sets, 
hence huge $ m $, for which Cholesky decomposition of the observation inter-correlation
$ ( m \times m ) $-matrix $ \Sigi{d}^\rho $ is computationally unfeasible even for actual finite-range Gaussian RF models. 
Several approximation techniques based on blocking of the matrix are studied, 
see \cite{Chen2021}.
We recommend a mitigation strategy that enforce some type of localization. The localization is obtained by assuming, minimum $ k $-link influence. 
This assumption entails that each entry in the inverted $ ( m \times m ) $-matrix 
$ [ \Sigi{d}^\rho ]^{-1} $ is only influenced by observations located 
less than $ k \tau_0 $ away. 
Recall that for the spatial Kernel predictor under the finite-range Gaussian RF assumptions, the observation at location $ \vect{x} $ has finite influence inside a ball with radius $ \tau_0 $ centered at location $ \vect{x} $. Hence there must be a minimum of $ k $ transitions through observation locations to reach the distance $ k \tau_0 $, and observation locations being excluded, must be at least $ k $-links away. The localization parameter $ k $ must be user-specified.

In the localized Kernel predictor the troublesome inverse observation 
inter-correlation $ ( m \times m ) $-matrix $[ \Sigi{d}^\rho]^{-1} $   is replaced by an approximation  $ [ \Sigi{d}^\rho ]^{-1 \ast} $. 
The approximation is based on the inversion of localized observation inter-correlation sub-matrices at each of the $ m $ observation locations. A combination of these sub-matrices  provides an approximation to the inverse of 
the full observation inter-correlation $ ( m \times m ) $-matrix.
The algorithm for obtaining this approximate matrix is,
\begin{algorithm}[Observation inter-correlation matrix approximation]
	\begin{align*}
		&\mbox{Initiate, } \\
		&\mbox{Localization range: } \Delta \in \mathbb{R}_+ \\
		&\mbox{Support matrix: } \matr{ \Psi } = 0 \Idm{m} 
		\mbox{ - dim } (m \times m ) \\
		&\mbox{For } \vect{x}_i^d ; i = 1, \dots , m \\
		&\mbox{Define neighborhood set: } 
		\texttt{M}_{\vect{x}_i^d }^\Delta 
		= \{ \vect{y} | \vect{y} \in \texttt{M} , | \vect{x}_i^d - \vect{y}| < \Delta \}
		\mbox{ - dim }  m_{ \vect{x}_i^d }^\Delta  \\
		&\mbox{Construct matrix: } 
		\Sigi{d_{ \vect{x}_i^d }^\Delta }^\rho 
		= \mbox{ Sub-matrix } \{ \Sigi{d}^\rho ; \texttt{M}_{\vect{x}_i^d }^\Delta \} 
		\mbox{ - dim } ( m_{ \vect{x}_i^d }^\Delta \times m_{ \vect{x}_i^d }^\Delta )  \\
		&\mbox{Compute: } 
		[ \Sigi{d_{ \vect{ x}_i^d}^\Delta }^\rho]^{-1} \mbox{ - dim }
		( m_{ \vect{x}_i^d }^\Delta \times m_{ \vect{x}_i^d }^\Delta )  \\
		&\mbox{Copy: }  
		[ \matr{ \Psi } ]_{ij} 
		= [ [\Sigi{d_{ \vect{ x}_i^d}^\Delta}^\rho]^{-1} ]_{\vect{x}_i^d \vect{y} } 
		\mbox{ - for corresponding } j \mbox{ and }  \vect{y} \in \texttt{M}_{\vect{x}_i^d}^\Delta \mbox{ entries } \\
		&\mbox{End For }  \\
		&\mbox{Define: } 
		[ \Sigi{d}^\rho ]^{ -1 \ast} = 1/2 \times [ \matr{ \Psi} + \matr{ \Psi}^T ]
	\end{align*}
\end{algorithm}
The observation inter-correlation $ ( m \times m ) $-matrix $ \Sigi{d}^\rho $ has lines and columns corresponding to the entries in the observation location set $ \texttt{M} $. 
The operator $  \mbox{Sub-matrix}  \{ \Sigi{d}^\rho ; \texttt{M}_{\vect{x}_i^d}^\Delta \} $ delivers the sub-matrix of $ \Sigi{d}^\rho $ containing the line and columns corresponding to the observation location sub-set $ \texttt{M}_{\vect{x}_i^d}^\Delta \subset \texttt{M} $, hence removing the line and columns corresponding to $ \texttt{M} \setminus \texttt{M}_{\vect{x}_i^d}^\Delta $.
The support $ ( m \times m) $-matrix $ \matr{ \Psi } $ will be sparse since 
the location set $ \texttt{M}_{ \vect{x}_i^d } $ is a sub-set of $ \texttt{M} $,
hence not all entries in line $ i $ will be assigned new values. 
Also the approximation $ [\Sigi{d}^\rho]^{ -1 \ast} $ will be sparse and it will be symmetric, but usually not non-negative definite.

The corresponding localized Kernel predictor is defined with localization range 
$ \Delta = k \tau_0 $ and it is,
\begin{align}  \label{eq:Loc-Func-pred}
	\{ \hat{r}^\ast ( \vect{x} )
	&= \mu_r^\ast + \vect{ \rho }_{xd}  \vect{ \alpha}^{d \ast} 
	= \mu_r^\ast + [ \vect{ \nu } ( \vect{x} ) ]^T \vect{ \alpha}^{d \ast} \\
	&= \mu_r^\ast + \Sigma_{ \vect{y} \in \texttt{M} } 
	\alpha_{ \vect{y} }^{d \ast} \nu_{ \vect{y} } ( \vect{x} )  \nonumber \\
	&= \mu_r^\ast + \Sigma_{ \vect{y} \in \texttt{M}_x^{\tau_0 } } 
	\alpha_{ \vect{y} }^{d \ast} \nu_{ \vect{y} } ( \vect{x} )
	; \vect{x} \in \texttt{D}  \} \nonumber
\end{align}
with the approximate weight $ m $-vector 
$ \vect{ \alpha}^{ d \ast } 
= [ \Sigi{d}^\rho ]^{-1 \ast} ( \vect{r}^d - \mu_r^\ast \iv{m} ) $. The associated localized prediction variance is,
\begin{align}  \label{eq:Loc-Func-pred-var}
	\{  \sigma_p^{2 \ast} ( \vect{x} ) &= \sigma_r^{ 2 \ast } 
	[ 1 -\vect{ \rho}_{xd} [ \Sigi{d}^\rho ]^{-1 \ast}  \vect{ \rho}_{dx} ]   
	=  \sigma_r^{ 2 \ast } [ 1 - [ \vect{ \nu} ( \vect{ x} ) ]^T  [ \Sigi{d}^\rho ]^{-1 \ast} 
	\vect{ \nu} ( \vect{ x} ) ]  \\
	&= \sigma_r^{ 2 \ast }
	[ 1 -  \Sigma_{ \vect{y}' \in \texttt{M} } \Sigma_{ \vect{y}'' \in \texttt{M} } 
	\beta_{\vect{y}' \vect{y}'' }^\ast
	\nu_{ \vect{y}' } ( \vect{x}  ) \nu_{ \vect{y}'' } ( \vect{x} ) ]    \nonumber \\
	&= \sigma_r^{ 2 \ast } 
	[ 1 -  \Sigma_{ \vect{y}' \in \texttt{M}_{ \vect{x} }^{ \tau_0} }
	\Sigma_{ \vect{y}'' \in \texttt{M}_{ \vect{x} }^{  \tau_0}  } 
	\beta_{\vect{y}' \vect{y}'' }^\ast
	\nu_{ \vect{y}' } ( \vect{x} ) \nu_{ \vect{y}'' } ( \vect{x} ) ]  
	; \vect{x} \in \texttt{D} \} \nonumber
\end{align}
with $ \beta_{\vect{y}' \vect{y}'' }^\ast 
= [[ \Sigi{ d }^\rho ]^{-1 \ast} ]_{ ij } $ ;
where $ ( i,j) $ and 
$ ( \vect{y}' , \vect{y}'' ) $ are matching observation numbers and locations.
The approximate prediction variances are not ensured to be in the eligible interval
$ [ 0.0, \sigma_r^{ 2 \ast } ] $, hence post-adaption must be made.
Both predictions and prediction variances are on a functional representation without any spatial grid discretization. The last equality in the expressions above follows from the spatial correlation function having finite range, 
$ \rho_r ( \vect{\tau} ) = 0.0 ; | \vect{\tau } | > \tau_0 $, and it causes the expressions to be spatially local which robustifies the predictor against approximations by localization.

The model parameters $ [ \mu_r , \sigi{r} ] $, being the expectation and variance levels respectively, can be estimated from Expression \ref{eq:MLE-ExpVar} as,
\begin{align*}
	\mu_r^\ast  
	&= [ \iv{m}^T [ \Sigi{d}^\rho ]^{-1 \ast} \iv{m} ]^{-1} 
	\times \iv{m}^T [ \Sigi{d}^\rho  ]^{-1 \ast} \vect{r}^d \\
	\sigma_r^{2 \ast} 
	&= m^{-1} ( \vect{r}^d - \mu_r^\ast \iv{m} )^T [ \Sigi{d}^\rho ]^{-1 \ast} 
	( \vect{r}^d - \mu_r^\ast \iv{m} ),   
\end{align*}
hence they can easily be assessed.

The computational demands are extremely favorable. The processing time is proportional to the number of observations $ m $, almost independent of the dimension of the reference space $ \texttt{D} $. Since no spatial discretization is involved the processing time is of course independent of any grid geometry. Moreover the approximation algorithm is very simple to implement and it is suitable for parallell processing on a computer.

The localized Kernel predictor appears with correct spatial characteristics according to the prior Gaussian RF model, the approximation will cause no spatial discontinuities. The localized predictor will however be non-exact in the sense that exactly observed values will not be exactly reproduced by the predictor. The deviation between exact observations and predictions in observation locations can provide a quantification of the approximation by calculating the deviation variance $ \sigi{ \ast } $.
The deviations will be zero for the global Kernel predictor and increase with decreasing $ k $ and rougher approximation. 
The prediction variance can be adjusted for this approximation uncertainty, and a reliable expression is 
$ \{ \hat{ \sigma}_p^2 ( \vect{x} )  = \sigma_p^{2 \ast} ( \vect{x} ) + \sigi{ \ast } 
; \vect{ x } \in \texttt{D} \} $, which very likely will take values in the eligible domain for variances.

\vspace{5 mm}
\noindent
\textbf{Summary }

To summarize - for moderate to large observation sets, hence $ m $ moderate to large, the observation inter-correlation $ ( m \times m) $-matrix $ \Sigi{d}^\rho $ may be sparsified by either assuming a finite-range Gaussian RF model or by using numerical tapering.
These approaches are equivalent, see Section \ref{sec:Corr-Func}, although a correct model formulation is to be prefered to a numerical trick by tapering. For moderate to large $ m $, the global Kriging and global Kernel predictors are available by 
one Cholesky decomposition of the sparse $ ( m \times m) $-matix $ \Sigi{d}^\rho $. The spatial Kernel predictor is on functional representation, and it should be prefered to the grid discretized representation provided by the spatial Kriging predictor. The former is actually the grid infill asymptotic limit of the latter.

For huge observation sets where not even sparsification of $ \Sigi{d}^\rho $ makes 
the Cholesky decomposition of
the matrix computationally feasible, localization approximations must be used for both predictors. The Kriging predictor is often simplified by assuming some sort of Markov property in the Gaussian RF model. The localized Kriging predictor requires a reduced observation inter-correlation matrix to be Cholesky decomposed for each 
grid node $ \vect{x} \in \texttt{L} \subset \texttt{D} $, 
hence $ n $ times. The Kernel predictor can be dimension reduced by assuming - minimum  
$ k $-link influence - in the Gaussian RF model. The localized Kernel predictor requires a reduced observation inter-correlation matrix to be Cholesky decomposed for each observation location 
$ \vect{x} \in \texttt{M} $, hence $ m $ times. Note that usually $ n \gg m $, hence one will expect that the localized Kernel predictor is far more computationally efficient than the localized Kriging predictor. Moreover, since the spatial Kernel predictor is on a functional representation, it should be preferred to the spatial Kriging prediction which is on a grid discretized representation.

\vspace{5 mm}
\noindent
\textbf{Example B }

This example demonstrates the characteristics of the localized Kernel predictor on a real observation set with some challenging features. No thorough empirical evaluation nor any comparison with competing predictors are performed. Empirical evaluations and comparisons are left for future studies.

The spatial variable under study is accumulated precipitation in millimeters during January-December 1997 in a sub-region of the US.  The observation set consists of $ 1330 $ locations with observed precipitation values, after removing missing and exact-zero observations. The full set of observed monthly precipitation for US 1895–1997 \citep{Nychka23} has its origin from US National Climatic Data Center and  \cite{Johns2003} contains a thorough discussion. 
The observation set is not really huge, but it is easily tractable and sufficiently large to demonstrate the characteristics of the localized Kernel predictor.

\begin{figure}  
	\centering
	\begin{subfigure}[ht]{0.3\textwidth}
		\includegraphics[width=1.0\textwidth]{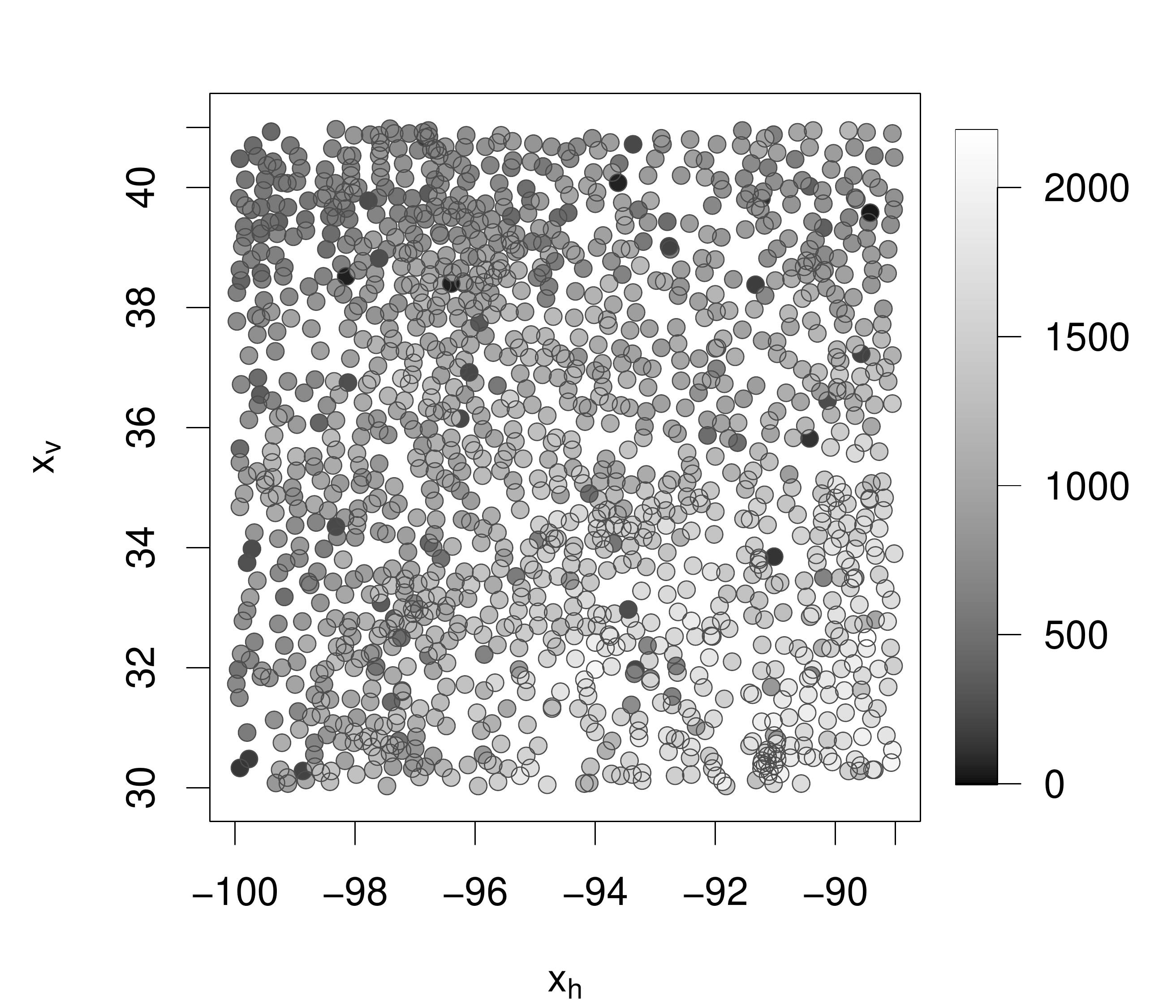}
		\caption{Locations and values - 1330 observations.}
	\end{subfigure}
	\begin{subfigure}[ht]{0.3\textwidth}
		\includegraphics[width=1.0\textwidth]{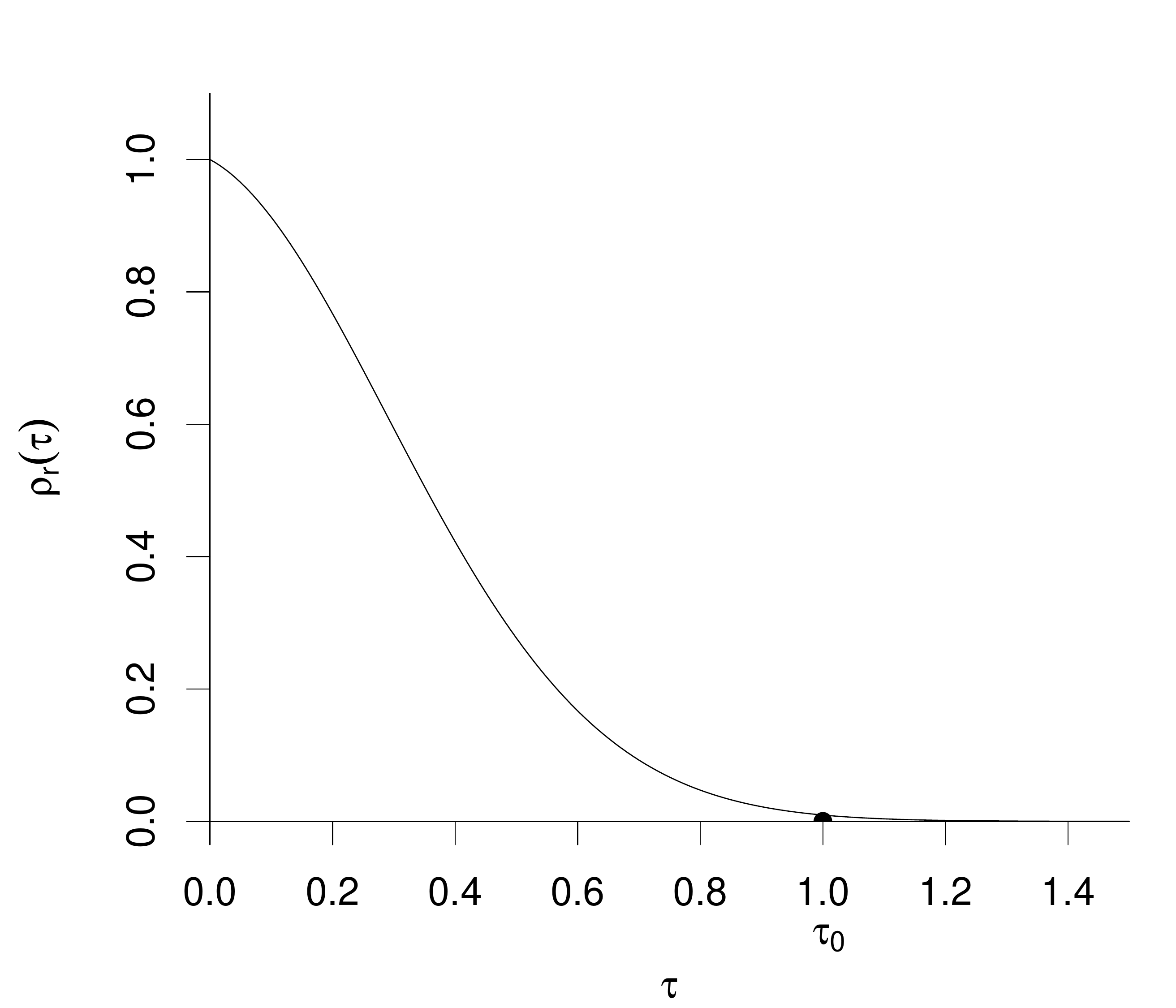}
		\caption{Finite-range spatial correlation function.}
	\end{subfigure}
	\caption{Observation set and model parameters.} 
	\label{fig:Obs-set-mod}
\end{figure}

Figure \ref{fig:Obs-set-mod} displays the $ m=1330  $ observation locations with exactly observed values of precipitation. 
The spatial reference $ \vect{x} = ( x_h, x_v ) \in \texttt{D} \subset \mathbb{R}^2 $ is defined to be $ ( \mbox{longitude,latitude} ) $, and the distance between two locations is defined to be the Euclidean distance.
The observations are relatively evenly located in the spatial reference domain. The values, assumed to be exact observations of the precipitations, seem to have a slight trend in the north-west direction. Spread across the domain several single extreme values are located in areas with fairly homogeneous values. These spike-patterns are difficult to capture and reproduce in the predictions.
In spite of the slight trend, a stationary, isotropic prior Gaussian RF model is defined. The model parameters $ [ \mu_r , \sigi{r} , \vect{ \eta }_r ] $ are the expectation and variance levels and the parameters of 
the spatial correlation function $ \rho_r ( \vect{ \tau } ; \vect{ \eta }_r ) $. 
After a short preliminary study we assign a finite-range second-order exponential correlation function, see Section \ref{sec:Corr-Func},
\begin{align*}
	\rho_r ( \vect{ \tau} ) =
	\rho_B ( \vect{ \tau }  ) \times \rho_R ( \vect{ \tau}  ) 
	=\left\{
	\begin{array}{lll} 
		1.0   &   | \vect{ \tau } | = 0.0 \\
		\exp \{ - ( | \vect{ \tau } | / 0.5 )^2 \}      \times
		(1 +  | \vect{ \tau } | / 2    ) (1 - | \vect{ \tau } |  )^2  & 0.0 < | \vect{ \tau} | < 1.0 \\
		0.0  &  | \vect{ \tau } | \geq 1.0
	\end{array}
	\right.
\end{align*}
The correlation function is displayed in Figure \ref{fig:Obs-set-mod}, and the finite range is $ \tau_0 = 1.0 $. It will typically be $ 30 $ observations from the set located within the range radius $ \tau_0  $.
The two other model parameters $ [ \mu_r , \sigi{r} | \vect{ \eta }_r ] $ will be estimated from the observations later. The observation inter-correlation $ ( m \times m ) $-matrix
$ \Sigi{d}^\rho $ contains 
entries $ [ \Sigi{d}^\rho ]_{ \vect{y}' \vect{y}''} = \rho_r ( \vect{y}' - \vect{y}'' ) 
; \vect{y}' , \vect{y}'' \in \texttt{M}  $.
The corresponding observation kernel function $ m $-vector 
$ \{ \vect{ \nu } ( \vect{x} ) ; \vect{x} \in \texttt{D} \} $   
contains entries $ \nu_{ \vect{y} }  ( \vect{x} ) 
= \rho_r ( \vect{x} - \vect{y} ) ; \vect{y} \in \texttt{M}  $. 
Since the spatial correlation function is defined to have finite range, both the matrix and the vector will be sparse.

The localized Kernel predictor, as defined above, relies on the specification of a neighborhood parameter $ k $. After a short preliminary study with inspection of the deviations between the values and their predictions in the observation locations we decide that $ k= 2 $ provides reliable results. Hence the neighborhood has 
radius $ k \tau_0 = 2.0 $, and it typically contains $ 120 $ observations from the set, being four times higher than the number of observations within the range radius. The computational demand typically includes 1330 Cholesky decompositions of $ ( 120 \times 120 ) $-matrices which all are sparse. The approximate inverse observation inter-correlation $ ( 1330 \times 1330 ) $-matrix 
$ [ \Sigi{d}^\rho ]^{-1 \ast} $ is obtained. The deliveries from the localized Kernel predictor are prediction and prediction variance on functional representations over the study domain. The model parameters expectation and variance level are estimated to 
$ \mu_r^\ast = 1076 $ and $ \sigma_r^{ 2 \ast } = 190 303 $.

\begin{figure}  
	\centering
	\begin{subfigure}[ht]{0.3\textwidth}
		\includegraphics[width=1.0\textwidth]{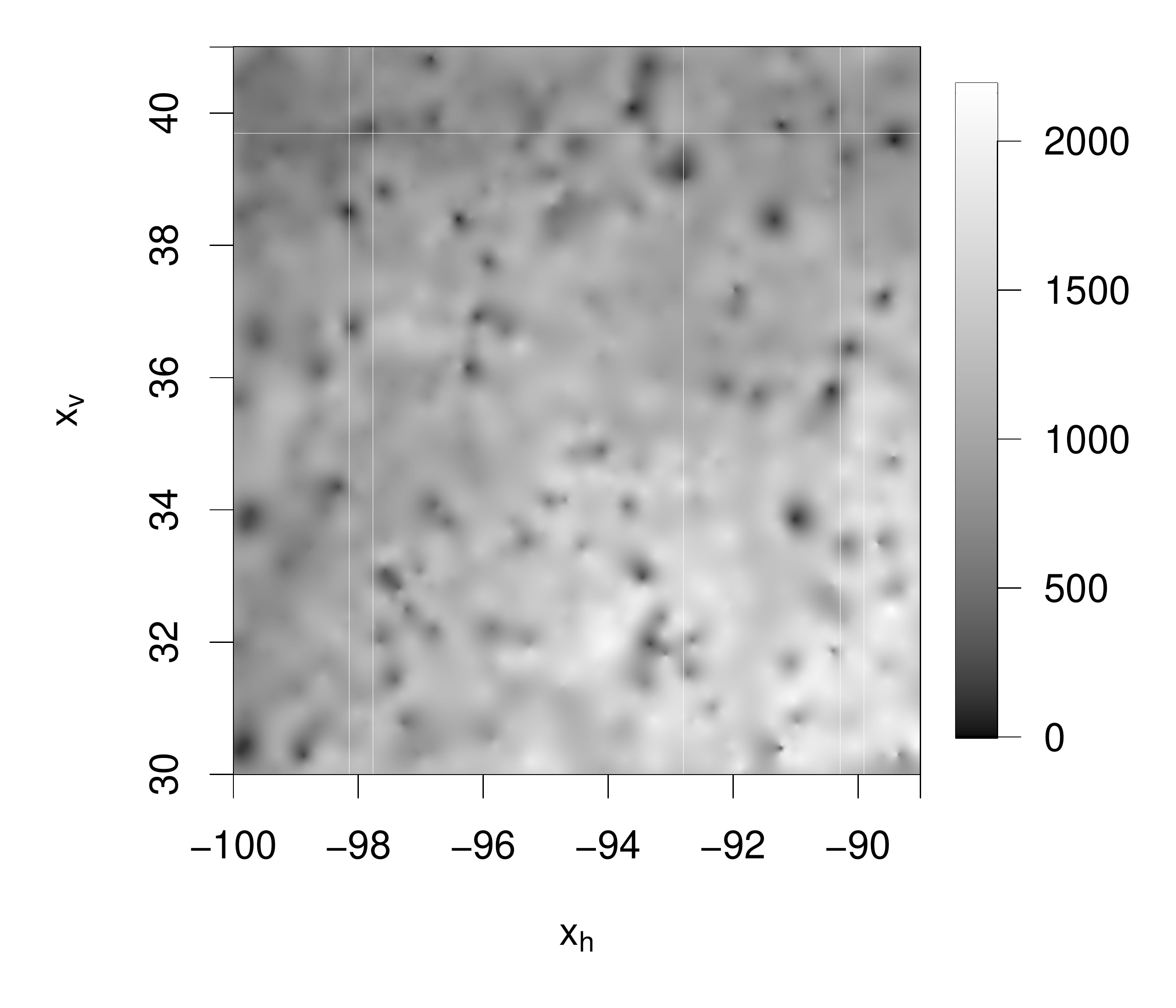}
		\caption{Prediction map.\newline}
	\end{subfigure}
	\begin{subfigure}[ht]{0.3\textwidth}
		\includegraphics[width=1.0\textwidth]{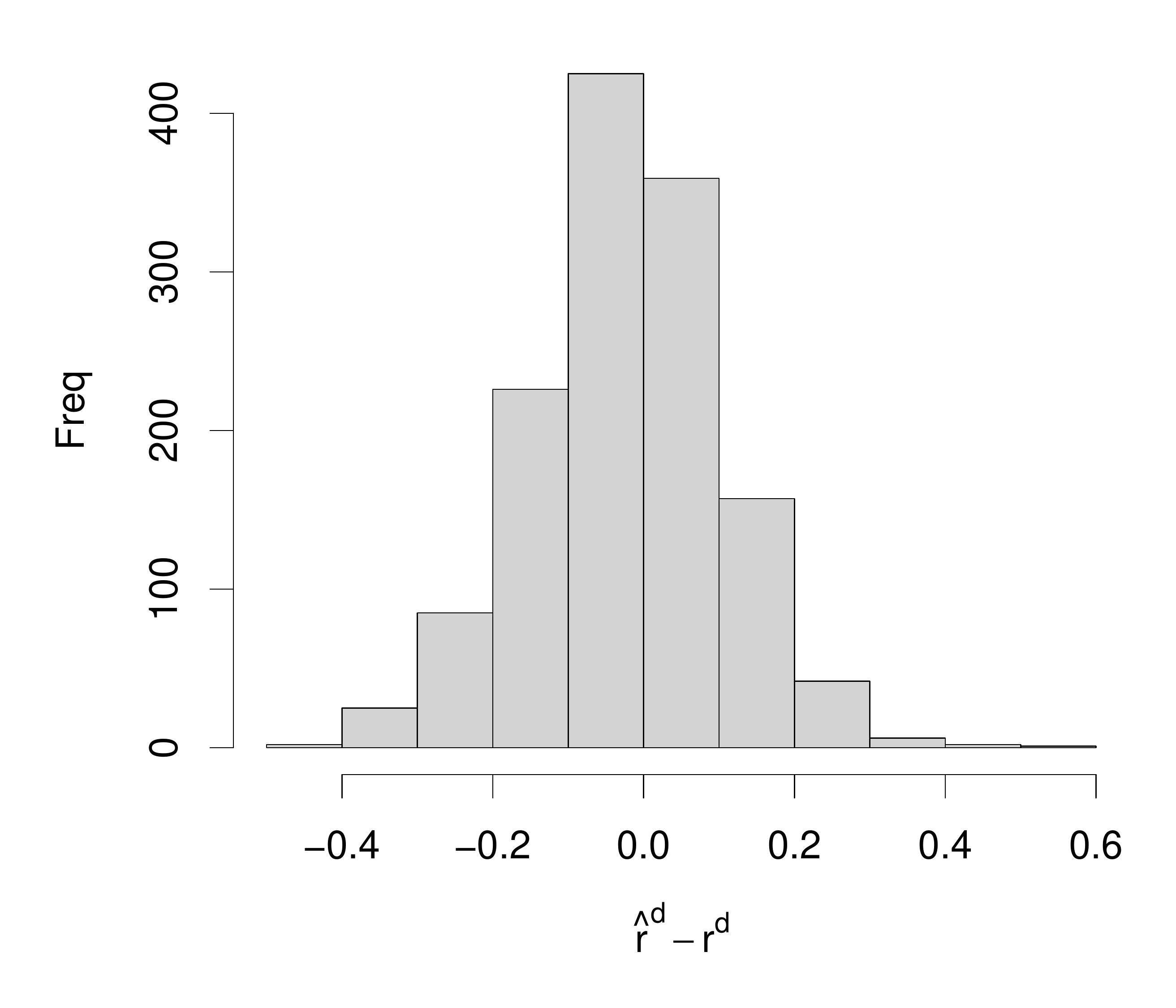}
		\caption{Prediction error histogram in observation locations.}
	\end{subfigure}
	\begin{subfigure}[ht]{0.3\textwidth}
		\includegraphics[width=1.0\textwidth]{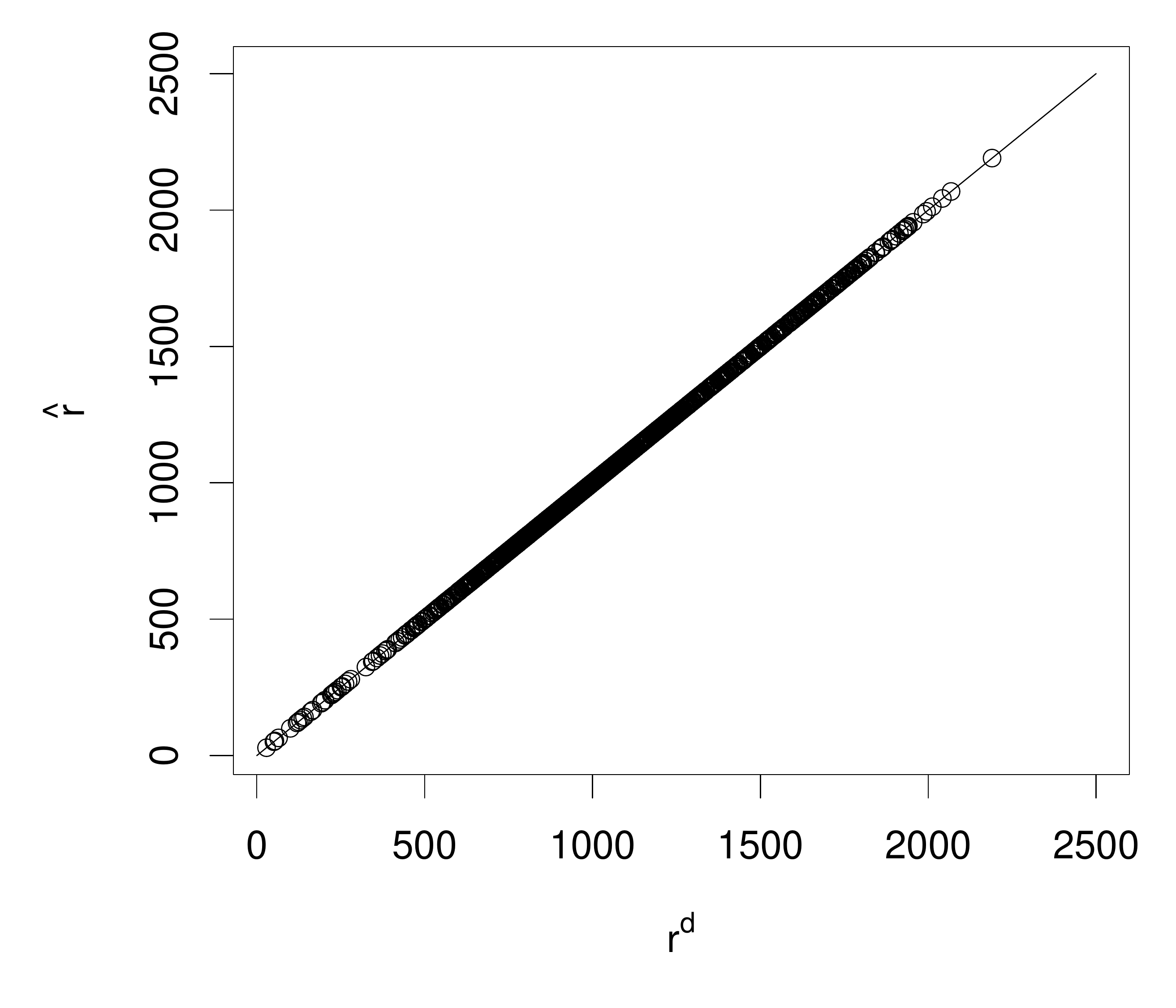}
		\caption{Value vs prediction cross-plot in observation locations.}
	\end{subfigure}
	\caption{Localized Kernel predictor: Neighborhood parameter $ k = 2 $. } 
	\label{fig:Loc-Func-Pred}
\end{figure}

\begin{figure}  
	\centering
	\begin{subfigure}[ht]{0.3\textwidth}
		\includegraphics[width=1.0\textwidth]{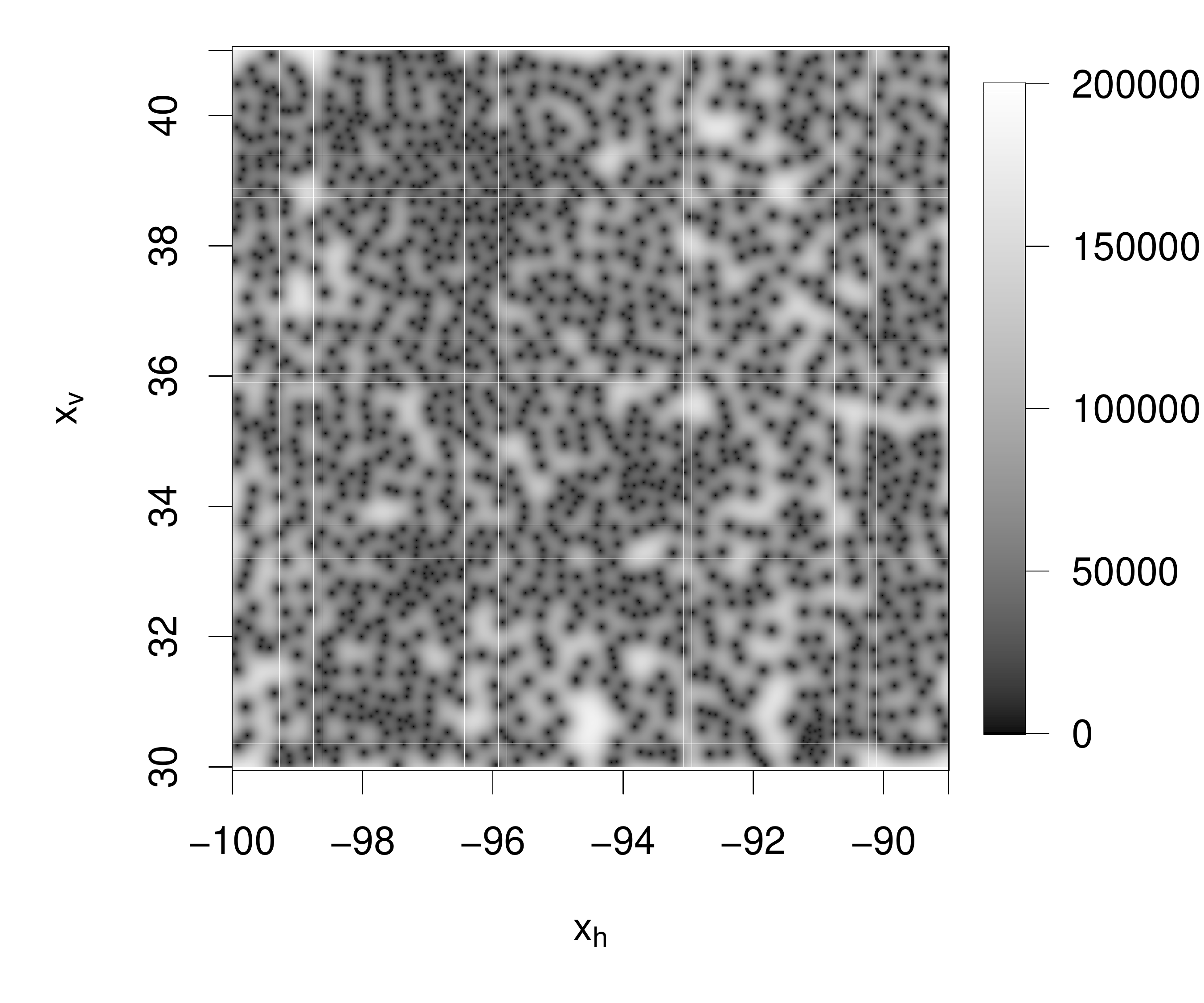}
		\caption{Prediction variance map. \newline}
	\end{subfigure}
	\begin{subfigure}[ht]{0.3\textwidth}
		\includegraphics[width=1.0\textwidth]{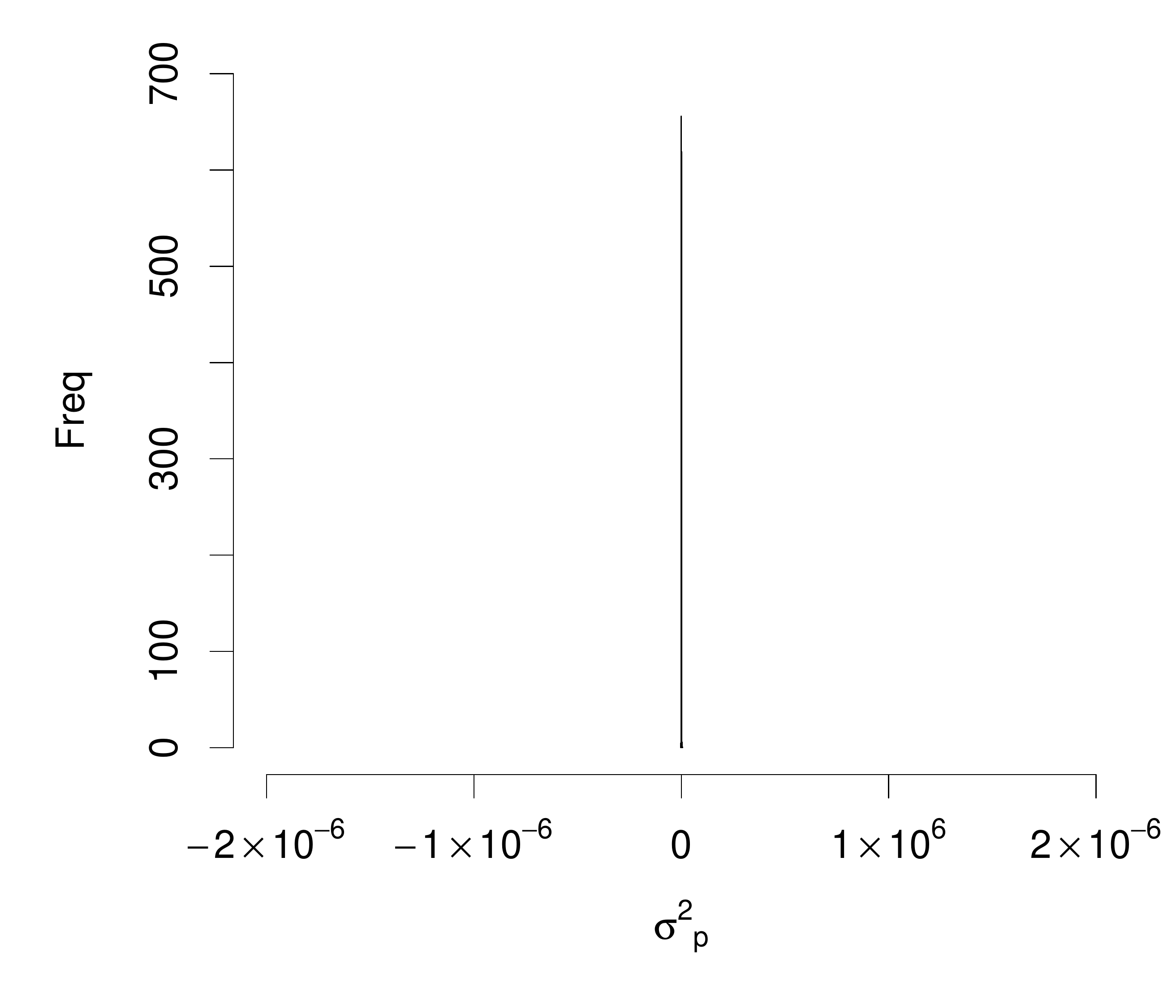}
		\caption{Prediction variance histogram in observation locations.}
	\end{subfigure}
	\caption{Localized Kernel predictor: Neigborhood parameter $ k = 2 $.  } 
	\label{fig:Loc-Func-Pred-Var}
\end{figure}

Figure \ref{fig:Loc-Func-Pred} contains three displays: the prediction map, a prediction error histogram and a prediction error cross-plot. The prediction is based on Expression 
\ref{eq:Loc-Func-pred}  which is on functional form, and it is transfered to lattice
resolution $ ( 1101 \times 1101 ) $ in the map. 
The spatial smoothness of the map is according to the Gaussian RF model without localization artifacts.
Note that the spike-patterns are reproduced. In the global Kernel predictor 
the observed values will be exactly reproduced but in the localized predictor they may not be.
The histogram in Figure \ref{fig:Loc-Func-Pred} displays the prediction errors in the observation locations, which appear as very small with 
deviation variance $ \sigi{ \ast} = 1.6 \times 10^{-2} $, almost ignorable compared to the variance of the observations. The observed values versus predictions are displayed as a cross-plot in the figure, and all pairs appear almost exactly at the unit diagonal as they should for exact prediction.
Figure \ref{fig:Loc-Func-Pred-Var} contains two displays: the prediction variance map and a prediction variance histogram. 
The prediction variance is based on Expression \ref{eq:Loc-Func-pred-var} which is on functional form, while the map has the same resolution as above. Note that the prediction variances approach zero at the observation locations, as they should. The histogram in 
Figure \ref{fig:Loc-Func-Pred-Var} displays the prediction variances at the observation location. This histogram will be a spike at zero for the global Kernel predictor, and it is almost a spike also for the localized approximation. 
Note that some negative values appear, which of course are not eligible as prediction variances. The prediction variances are in 
the range $ [ -2 \times 10^{-9} , 2 \times 10^{-9} ] $,   hence extremely small.  A more reliable expression for the prediction variance, also capturing the approximation uncertainty, is 
$ \{ \hat{ \sigma }_p^2 ( \vect{x} )= \sigma_p^{  2 \ast} ( \vect{x} ) + \sigi{ \ast } 
; \vect{x} \in \texttt{D} \} $, which will take values in the eligible range for variances. 

The results obtained in this example are very encouraging. The localized Kernel predictor seems to provide highly reliable predictions and prediction variances even with quite modest neighborhoods. We believe studies with really huge observation sets can be made with computational demands proportional to the number of observations, hence independent of any grid representation.

\section{Spatial Finite-range Correlation Functions} \label{sec:Corr-Func}

The basic references for spatial correlation functions are \cite{Gneiting1999} and \cite{Gneiting2002}. The discussion in this section is primarily based on the latter.

A spatial correlation function 
$ \rho ( \vect{ \tau } ) ; \mathbb{R}^q \rightarrow \mathbb{R} $ need to be a non-negative definite function, hence
\begin{align*}
	\Sigma_{i=1}^n \Sigma_{j=1}^n 
	\beta_i \beta_j \rho ( \vect{ x}_i - \vect{x}_j ) \geq 0
\end{align*}
for all configurations 
$ \vect{x} = ( \vect{x}_1 , \dots , \vect{x}_n ) \in \mathbb{R}^{q \times n} $, for all
$ \vect{ \beta } = ( \beta_1, \dots , \beta_n ) \in \mathbb{R}^n $ and for all
$ n \in \mathbb{N}_+ $.
Moreover, the function value at the origin must be $ \rho ( 0 \iv{q} ) = 1.0 $.

It is simple to demonstrate that if both $ \rho_B ( \vect{ \tau } ) $ and
$ \rho_R ( \vect{\tau} ) $ are non-negative definite correlation functions, then both
\begin{align*}
	\rho_+ ( \vect{ \tau } ) 
	&= w \rho_B ( \vect{ \tau } ) + (1-w) \rho_R ( \vect{ \tau } )
	; \mathbb{R}^q \rightarrow \mathbb{R} \\
	\rho_\times ( \vect{ \tau } ) &= \rho_B ( \vect{ \tau } ) \times \rho_R ( \vect{ \tau } )
	; \mathbb{R}^q \rightarrow \mathbb{R} 
\end{align*}
for $ w \in \mathbb{R}_{[0,1]} $, belong to the same class of functions.

Consider a Gaussian RF 
$ \{ r ( \vect{x} ) ; \vect{x} \in \texttt{D} \subset \mathbb{R}^q \} $, with  reference space 
of dimension three or lower, hence $ q \leq 3 $. Denote $ \rho_B ( \vect{\tau} ) $ the spatial base correlation function, and let it be any valid parametric model, 
$ \rho_B ( \vect{\tau } ; \vect{ \eta}_B ) $. Denote  $ \rho_R ( \vect{\tau} ) $ the spatial finite-range correlation function, and let it in this study be the isotropic spherical model,
\begin{align*}
	\rho_R ( \vect{ \tau} ; \tau_0 ) =
	\left\{
	\begin{array}{lll} 
		1.0   &   | \vect{ \tau } | = 0.0 \\
		(1 + | \vect{ \tau } | / 2 \tau_0    ) (1 - | \vect{ \tau } | / \tau_0 )^2  
		& 0.0 < | \vect{ \tau} | < \tau_0 \\
		0.0  &  | \vect{ \tau } | \geq \tau_0
	\end{array}
	\right.
	; \mathbb{R}^q \rightarrow \mathbb{ R }
\end{align*}
Then the combined spatial correlation function
\begin{align*}
	\rho_{BR} ( \vect{ \tau} ; \vect{ \eta} ) &=
	\rho_B ( \vect{ \tau } ; \vect{ \eta}_B ) \times \rho_R ( \vect{ \tau} ; \tau_0 ) \\
	&=\left\{
	\begin{array}{lll} 
		1.0   &   | \vect{ \tau } | = 0.0 \\
		\rho_B ( \vect{ \tau} ; \vect{ \eta}_B ) \times
		(1 +  | \vect{ \tau } | / 2 \tau_0   ) (1 - | \vect{ \tau } | / \tau_0 )^2  
		& 0.0 < | \vect{ \tau} | < \tau_0 \\
		0.0  &  | \vect{ \tau } | \geq \tau_0
	\end{array}
	\right.
	; \mathbb{R}^q \rightarrow \mathbb{ R }
\end{align*}
will define a large class of spatial finite-range correlation functions parametrized by 
$ \vect{ \eta } = ( \vect{ \eta}_B , \tau_0 ) $.

Note that the corresponding sparse observation inter-correlation $ ( m \times m) $-matrix 
$ \Sigi{d}^\rho $ defined by $ \rho_{BR} ( \vect{ \tau }; \vect{ \eta} ) $ can be
expressed as $ [ \Sigi{d}^\rho ]_B \otimes \matr{T}_R $ with the former factor being defined by $ \rho_B ( \vect{ \tau } ; \vect{ \eta}_B ) $ and the latter sparse tapering matrix being defined by $ \rho_R ( \vect{ \tau } ; \tau_0 ) $. Hence assuming a finite-range Gaussian RF model and the numerical approximation using tapering produce the same results.
Philosophically, however, it is much easier to justify specification of a correct model than using numerical tricks.

\section{Conclusions}

The challenge is to assess the continuous spatial variable over a finite spatial domain,
based on observations in a set of locations. The usual approach is to generate a grid representation of the variable and to make simple spatial interpolation of the grid values to assess the value in an arbitrary non-grid location. The Kriging predictor is frequently used to provide the predictions and prediction variances in the grid nodes of the representation. For large observation sets the grid must be dense to provide a reliable representation, and in these cases both computer processing and storage can be challenging.

We define a spatial Kernel predictor which provides a functional representation of the prediction and the associate prediction variance. No grid discretization is involved. The computational and storage requirements are expected to be very modest since the predictor only is dependent on the number of observations.

Under Gaussian RF assumptions, the Kriging and Kernel predictors provide identical prediction and prediction variance for the spatial variable in an arbitrary location in the spatial domain. Both predictors are optimal in the square-error sense. The functional representation can be seen as the grid infill asymptotic limit of the grid representation.

For huge sets of observations the observation inter-correlation matrix may have dimensions such that Cholesky decomposition is computationally unfeasible. Both prediction and model parameter inference will for these cases appear as challenging. The usual mitigation strategy is to use approximate localized predictors and estimators. For Kriging based predictors a large variety of huge-data-approximations are defined and evaluated. Almost without exceptions these approximate prediction approaches rely on a grid representation 
and the grid density must be much higher than the observation density
to capture  the information in the observations.

We define a huge-data-approximation of the Kernel predictor, the localized Kernel predictor, which only depends on the number of observations. For studies with a huge set of observations, particularly for studies with three-dimensional spatial domains, and spatio-temporal studies, avoiding a grid discretization is very beneficial. Dramatic savings in computer demands, both processing and storage, is expected. The localization algorithm 
is also suitable for parallell processing on computers.
Moreover, model parameter inference of the expectation and variance levels can be very efficiently made.

The Kernel predictor is used in two examples in order to demonstrate the flexibility and potential of the methodology, and the results are very encouraging.  Comparison with other predictors on real cases with huge observation sets is of course of interest, but these comparisons are left for future studies.

\bibliographystyle{apalike}

\bibliography{ref}
\end{document}